\documentclass{JHEP3}

\usepackage{epsfig, multicol, bbm, amsmath, amssymb, euscript, array, amsfonts} 

\title{Unified Dark Matter scalar field models with fast transition}

\author{Daniele Bertacca$^{a,b,c}$, Marco Bruni$^{b}$, Oliver F. Piattella$^{d,e}$, Davide Pietrobon$^{f}$\\
$^a$ Dipartimento di Fisica Galileo Galilei  Universit\`{a} di Padova, via F. Marzolo, 8 I-35131 Padova, Italy\\
$^b$ Institute of Cosmology \& Gravitation, University of Portsmouth,
Dennis Sciama Building, Portsmouth, PO1 3FX, United Kingdom\\
$^c$ INFN Sezione di Padova, via F. Marzolo, 8 I-35131 Padova, Italy\\
$^d$ Department of Physics, Universidade Federal do Esp\'{i}rito Santo, avenida Ferrari 514, 29075-910, Vit\'{o}ria, ES, Brazil\\
$^e$ INFN, sezione di Milano, Via Celoria 16, 20133 Milano, Italy\\
$^f$ Jet Propulsion Laboratory, California Institute of Technology,  4800 Oak Grove Drive, 91109 Pasadena CA USA\\
E-mails: \email{daniele.bertacca@pd.infn.it, daniele.bertacca@port.ac.uk}, \email{marco.bruni@port.ac.uk}, \email{oliver.piattella@gmail.com}, \email{davide.pietrobon@jpl.nasa.gov} }

\preprint{\arXivid{10xx.xxxx}}
\abstract{\today}


\abstract{
We investigate the general properties of Unified Dark Matter (UDM)  scalar field models with Lagrangians with a non-canonical kinetic term, looking specifically for models
that  can produce a fast transition  between an early Einstein--de\! Sitter CDM-like era and a later  Dark Energy like phase,  similarly to the barotropic fluid  UDM models  in JCAP1001(2010)014. However, while the background evolution can be very similar in the two cases,  the  perturbations  are naturally adiabatic in fluid models, while in  the scalar field  case they are necessarily  non-adiabatic.
The new approach to building UDM Lagrangians proposed here allows  to escape the common problem of the fine-tuning of the parameters  which plague many UDM  models.
We analyse the properties of perturbations in our model, focusing on the the evolution of the effective speed of sound  and that of the Jeans length. With this insight, we can set theoretical constraints on the parameters of the model, predicting sufficient conditions for the model to be viable. An interesting feature of our models is that what can be interpreted as $w_{\rm DE}$ can be $<-1$ without violating the null energy conditions. 
}

\begin{document}

\section{Introduction}

In the last decade the $\Lambda$CDM model \cite{Peebles:1984ge, Efstathiou:1990xe} has emerged as the ``concordance'' model \cite{Spergel:2003cb, Tegmark:2003ud} of our Universe. It assumes General Relativity (GR) as the correct theory of gravity, and two unknown components dominating the late times dynamics: {\it i )} Cold Dark Matter (CDM), responsible for structure formation, {\it ii )} a cosmological constant $\Lambda$ making up the balance for a spatially flat Universe and driving the cosmic acceleration \cite{Perlmutter:1998np, Riess:1998cb, Riess:1998dv, Percival:2007yw, Percival:2009xn, Amanullah:2010vv}.

Issues (still open today) with $\Lambda$, viz. the so-called ``cosmological constant problem'' \cite{Weinberg:1989, Zlatev:1998tr}, prompted cosmologists to investigate alternatives to the cosmological constant, mainly in the form of a dynamic component dubbed Dark Energy (DE) \cite{Sahni:1999gb, Peebles:2002gy, Padmanabhan:2002ji, Copeland:2006wr, Tsujikawa:2010sc, Amendola:1272934}. By now, many independent observations support both the existence of a CDM component and that of a separate DE \cite{Perlmutter:1998np, Riess:1998cb, Riess:1998dv, Peacock:2006kj, Allen:2004cd, Tegmark:2006az, Percival:2006kh, Percival:2007yw, Percival:2009xn, Reid:2009xm, Amanullah:2010vv, Larson:2010gs, Komatsu:2010fb, Blanchard:2010gv}.

However, it should be recognised that, while some form of CDM is independently expected to exist within any modification of the Standard Model of high energy physics, the really compelling reason to postulate DE has been the acceleration in the cosmic expansion.

It is mainly for this reason that it is worth investigating the hypothesis that CDM and DE are two aspects of a single Unified Dark Matter (UDM) component, also referred to as ``Quartessence",  see e.g.\ \cite{Kamenshchik:2001cp, Bilic:2001cg, Bento:2002ps, Carturan:2002si, Sandvik:2002jz, Scherrer:2004au, Giannakis:2005kr, Bertacca:2007ux, Bertacca:2007cv, Bertacca:2007fc, Balbi:2007mz, Quercellini:2007ht, Pietrobon:2008js, Bertacca:2008uf, Bilic:2008yr, Camera:2009uz, Li:2009mf, Piattella:2009kt, Gao:2009me, Camera:2010wm, Lim:2010yk} (a more complete list of UDM models can be found in a recent review \cite{Bertacca:2010ct}).
In UDM models a single matter component  must drive both the accelerated expansion of the Universe at late times and the formation of structures. This poses an interesting challenge to model building with this single component. Indeed, the accelerated expansion in a feature of the homogeneous and isotropic background, while the description of the structure formation process requires to consider inhomogeneities, at least at the first perturbative order. This has to be contrasted with CDM+DE models, where the CDM is perturbed and drives structure formation, while in general DE  cannot cluster but drives the acceleration of the background. 

A large variety of UDM models have been investigated in the literature, mainly based on adiabatic fluids or on scalar field Lagrangians. For example, the generalised Chaplygin gas \cite{Kamenshchik:2001cp, Bilic:2001cg, Bento:2002ps, Sandvik:2002jz}, the Scherrer \cite{Scherrer:2004au} and generalised Scherrer solutions \cite{Bertacca:2007ux}, the perfect fluid with ``affine'' equation of state \cite{Balbi:2007mz, Pietrobon:2008js} (see \cite{Quercellini:2007ht} for the corresponding scalar field model), and the homogeneous scalar field deduced from the galactic halo space-time \cite{DiezTejedor:2006qh, Bertacca:2007fc}.

It is also possible to reinterpret UDM models based on a scalar field Lagrangian  with a non-canonical kinetic term in terms of non-adiabatic fluids \cite{Brown:1992kc, DiezTejedor:2005fz} (see also \cite{Bertacca:2007ux, Bertacca:2008uf, Bertacca:2010ct}). For these models, we can use Lagrangians with a non-canonical kinetic term  (namely a term which is an arbitrary function of the square of the time derivative of the scalar field, in the homogeneous and isotropic background).

Originally, these Lagrangians were proposed to obtain inflationary solutions driven by kinetic energy, the so-called $k$-inflation \cite{ArmendarizPicon:1999rj, Garriga:1999vw}. This scenario was also adopted as a description of DE \cite{Chiba:1999ka, dePutter:2007ny, Linder:2008ya}. When $k$-inflation was extended to embrace more general Lagrangians \cite{ArmendarizPicon:2000dh, ArmendarizPicon:2000ah}, the resulting new scenario was then dubbed $k$-essence (see also \cite{Vikman:2004dc, Rendall:2005fv, Babichev:2007dw, Linder:2008ya, Arroja:2010wy, Unnikrishnan:2010ag}).
An important feature of scalar field models is the effective speed of sound, which remains defined in the context of linear perturbation theory, is not the same as the adiabatic speed of sound (see \cite{Bardeen:1983qw, Bruni:1991kb, Garriga:1999vw, Mukhanov:2005sc}, cf.\ \cite{Hu:1998kj}).

Independently of the approach chosen to build a UDM model with a single matter component, in general the latter will have pressure perturbation in the rest frame (which is then gauge invariant), which implies a non vanishing  effective speed of sound (see  \cite{Hu:1998kj}, cf.\ also  \cite{Bardeen:1980kt, Kodama:1985bj, Bardeen:1983qw, Bruni:1991kb}). The latter corresponds to a Jeans length (i.e.\ a sound horizon) below which clustering is suppressed \cite{Hu:1998kj, Bertacca:2007cv, Pietrobon:2008js}. Moreover, the evolution of the gravitational potential on scales smaller than the Jeans length is characterised by oscillations and decay, thus producing a strong late-times Integrated Sachs Wolfe (ISW) effect which may spoil the agreement with the observed Cosmic Microwave Background (CMB) radiation angular power spectrum \cite{Bertacca:2007cv}. Therefore, the general lesson to be drawn is that a viable UDM model must necessarily be characterised by a vanishingly small effective speed of sound, so that structure formation would not be suppressed (at least in the linear regime) and the ISW effect would be compatible with CMB observation.

However, for many UDM models the requirement of a vanishingly small speed of sound implies such a severe fine-tuning on the parameters that they become practically indistinguishable from $\Lambda$CDM, see for example \cite{Sandvik:2002jz, Scherrer:2004au, Giannakis:2005kr, Pietrobon:2008js, Piattella:2009da}. Fortunately, this does not need to be the case in general: indeed, in order to avoid such undesirable feature, some possible way-outs have recently been proposed.
In \cite{Piattella:2009kt} we introduced a class of UDM models whose equation of state allows for a fast transition between an early CDM-like era and a later $\Lambda$CDM-like phase. The speed of sound can be very large during the transition, but if the latter is fast enough, these models predict a large-scale structure spectrum and an angular CMB temperature spectrum in agreement with observation. Ref.~\cite{Gao:2009me} explored unification of DM and DE in a theory containing a scalar field of non-Lagrangian type, obtained by direct insertion of a kinetic term into the energy-momentum tensor. Ref.~\cite{Lim:2010yk} introduced a class of models based on two scalar fields, one of which is a Lagrange multiplier enforcing a constraint between the other scalar field and its derivative. The purpose is to assure a vanishing speed of sound. In \cite{Bertacca:2008uf}, the authors devised a reconstruction technique for Lagrangian which allows to find models where the effective speed of sound is so small that the {\it k}-essence scalar field can cluster (see also \cite{Camera:2009uz, Camera:2010wm, Bertacca:2010ct}).

In the present paper we develop the technique of \cite{Bertacca:2008uf} to build scalar field Lagrangians with non-canonical kinetic term which can produce a fast transition,  similarly to the barotropic fluid  UDM models  we proposed in \cite{Piattella:2009kt}. However, while the background evolution can be very similar in the two cases,  the  perturbations  are naturally adiabatic in fluid models, while in  the scalar field  case they are necessarily  non-adiabatic \cite{Bardeen:1983qw,Bruni:1991kb, Mukhanov:2005sc}, cf.\ \cite{Hu:1998kj, Bardeen:1980kt, Kodama:1985bj}.
This new approach allows us to escape the problem of the fine-tuning on the parameters which usually plague many UDM models.

The rest of the paper is organised as follows. In Section~\ref{sec:bgperteq} we present the basic equations describing the background and the perturbative evolution of a general $k$-essence Lagrangian. In Section~\ref{prescription} we generalise the class of UDM models investigated in \cite{Bertacca:2008uf} and in Section~\ref{p-fast} we introduce a  UDM model with fast transition, analysing its background evolution. In Section~\ref{sec:perts} we analyse the properties of perturbations in this model, focusing on the the evolution of the effective speed of sound, of the Jeans length and of the gravitational potential during the transition. Section \ref{sec:concl} is devoted to our conclusions.\\
Throughout the paper we use $8\pi G = c^2 = 1$ units and the $(-,+,+,+)$ signature for the metric. Greek indices run over $\{0,1,2,3\}$, denoting space-time coordinates, whereas Latin indices run over $\{1,2,3\}$, labelling spatial coordinates.

\section{Background and perturbative equations for UDM scalar field models}\label{sec:bgperteq}

\label{basics}
The action describing a generic Unified Dark Matter scalar field model in GR can be written as
\begin{equation}\label{eq:action}
S = S_{G} + S_{\varphi}= 
\int d^4 x \sqrt{-g} \left[\frac{R}{2}+\mathcal{L}(\varphi,X)\right] \, ,
\end{equation}
where 
\begin{equation}\label{x}
X=-\frac{1}{2}\nabla_\mu \varphi \nabla^\mu \varphi \;.
\end{equation}
The energy-momentum tensor of the scalar field $\varphi$ is
\begin{equation}
  \label{energy-momentum-tensor}
  T^{\varphi}_{\mu \nu } = 
- \frac{2}{\sqrt{-g}}\frac{\delta S_{\varphi }}{\delta
    g^{\mu \nu }}=\frac{\partial \mathcal{L}
(\varphi ,X)}{\partial X}\nabla _{\mu }\varphi
  \nabla _{\nu }\varphi +\mathcal{L}(\varphi ,X)g_{\mu \nu }.
\end{equation}
If $\nabla _{\mu }\varphi$ is time-like, $S_{\varphi}$ describes a perfect fluid with 
$T^{\varphi}_{\mu \nu }=(\rho +p) u_{\mu} 
u_{\nu }+p\,g_{\mu \nu }$, with pressure 
\begin{equation}
  \label{pressure}
  \mathcal{L}=p(\varphi ,X)\;,  
\end{equation}
and energy density
\begin{equation}
  \label{energy-density}
  \rho =\rho 
(\varphi ,X)= 2X\frac{\partial p(\varphi ,X)}
  {\partial X}-p(\varphi ,X)\;,
\end{equation}
where 
\begin{equation}
  \label{eq:four-velocity}
  u_{\mu }= \frac{\nabla _{\mu }\varphi }{\sqrt{2X}}
\end{equation}
is the ``energy frame" four-velocity \cite{Bruni:1992dg, Bruni:1991kb}. In this frame, the kinetic term becomes $X=\dot{\varphi}^2/2$, where $\dot{\varphi}= u^{\mu} \nabla _{\mu}$ is the proper time derivative along $u^\mu$.

Now we assume a flat Friedmann-Lema\^{i}tre-Robertson-Walker (FLRW) background metric with scale factor $a(t)$. When the energy density of radiation becomes negligible, and disregarding also the small baryonic component, the background evolution of the Universe is completely described by the following equations: 
\begin{eqnarray}
\label{eq_u1}
H^2 &=& \frac{1}{3}  \rho\, , \\
\label{eq_u2}
\dot{H} &=& - \frac{1}{2} (p + \rho)\, ,
\end{eqnarray}
where the dot denotes differentiation with respect to the cosmic time $t$ (coinciding with proper time along $u^\mu$) and $H=\dot{a}/a$ is the Hubble parameter. Eqs.\ (\ref{eq_u1})-(\ref{eq_u2}) imply the energy conservation equation 
\begin{equation}
\label{coneq}
 \dot{\rho} = -3H\left(\rho + p\right)\;.
\end{equation}

On the background FLRW metric, the equation of motion for the scalar field $\varphi(t)$ follows from the energy momentum tensor (\ref{energy-momentum-tensor}) and Eq.\ (\ref{coneq}):
\begin{equation}
 \label{eq_phi}
\left(\frac{\partial p}{\partial X} 
+2X\frac{\partial^2 p}{\partial X^2}\right)\ddot\varphi
+\frac{\partial p}{\partial X}(3H\dot\varphi)+
\frac{\partial^2 p}{\partial \varphi \partial X}\dot\varphi^2
-\frac{\partial p}{\partial \varphi}=0 \;. 
\end{equation}
The two relevant quantities that characterize the background and perturbative dynamics of a UDM model are the  equation of state $w \equiv p/\rho$ and the effective speed of sound $c_{\rm s}$, which, in our case, read 
\begin{eqnarray}
\label{w}
w &=& \frac{p}{2X (\partial p/\partial X) - p} \;, \\
\label{csUDMscalar}
c_{\rm s}^2 &\equiv&  \frac{\partial p /\partial X}
{\partial \rho /\partial X} = 
\frac{\partial p/\partial X}{(\partial p/\partial X)+ 2X(\partial^2 p/\partial X^2)} \;. 
\end{eqnarray}
The effective speed of sound plays a major role in the evolution of the scalar field
perturbations $\delta\varphi$ and in the growth of the 
overdensities $\delta\rho$.
In fact, let us consider small inhomogeneities of the scalar field, $\varphi(t,x)=\varphi_0(t)+\delta\varphi(t,\mbox{\boldmath $x$})$, and
write FLRW metric in the longitudinal gauge:
\begin{equation}
ds^2 = - (1+2 \Phi)dt^2 + a^2(t)(1-2 \Phi) \delta_{ij} dx^i dx^j \;,
\end{equation}
where $\delta_{ij}$ is the Kronecker symbol and we have used the fact that $\delta T_{i}^j = 0$ 
for $i \neq j$~\cite{Mukhanov:1990me}. Indeed, for this perfect fluid case, there is a unique peculiar gravitational potential $\Phi$, 
analogous to the Newtonian potential \cite{Bruni:1992dg}.
 When we linearize the $(0-0)$ and $(0-i)$ components of Einstein equations 
(see Refs.~\cite{Garriga:1999vw, Mukhanov:2005sc}), we obtain
the following second order differential equation \cite{Mukhanov:2005sc, Bertacca:2007cv}:
\begin{equation}
\label{diff-eq_u}
u''-c_s^2 \nabla^2 u - \frac{\theta''}{\theta}u=0\;
\end{equation}
where $u\equiv 2 \Phi/(p+\rho)^{1/2}$, $\theta \equiv (1+p/\rho)^{-1/2}/(\sqrt{3}a)$ and the prime denotes derivation w.r.t. the conformal time $\eta$, which is defined through $d\eta = dt/a$. As it is clear from \eqref{eq_phi} and \eqref{diff-eq_u}, the effective speed of sound plays a crucial role in the evolution of the scalar field perturbations $\delta\varphi$ and of the gravitational potential.

\section{Prescriptions for building a scalar field UDM model}\label{prescription}

A well-known issue of UDM models is the existence of an effective sound speed [see Eq.~(\ref{diff-eq_u})], which may become significantly different from zero during the Universe evolution and thus leading to the appearance of a Jeans length (i.e.\ a sound horizon) below which the dark fluid cannot cluster (e.g.\ see \cite{Hu:1998kj, Bertacca:2007cv, Pietrobon:2008js}). Indeed, the presence of a non-vanishing speed of sound modifies the evolution of the gravitational potential, producing a strong Integrated Sachs Wolfe (ISW) effect \cite{Bertacca:2007cv}. 
For this reason, one of the main issues in the framework of UDM model building is to investigate whether the single dark fluid is able to cluster and produce the cosmic structures we observe. 

In Ref.\ \cite{Bertacca:2008uf} the authors proposed a technique to construct UDM models where the scalar field mimics the $\Lambda$CDM background evolution and, at the same time, has an effective sound speed  that is small enough to allow for structure formation,  also avoiding a strong ISW effect (see also \cite{Camera:2009uz, Camera:2010wm}).

In this Section we develop and generalize the approach of Ref.\ \cite{Bertacca:2008uf}. Specifically, we focus on scalar field Lagrangians with non-canonical kinetic term to obtain Unified Dark Matter models where a single component can  mimic the dynamical effects of Dark Matter and Dark Energy. This sets the ground for Section \ref{p-fast}, where 
our aim is to build scalar field models in which the dynamics is characterised by a fast transition between an early CDM-like phase, with the background following an  Einstein de Sitter evolution, and a late accelerated DE-like phase, 
in this way  generalising the UDM fluid models with fast transitions we presented in \cite{Piattella:2009da}. 

We look for a scalar field Lagrangian $\mathcal{L}$ whose classical trajectories are directly described by an appropriate pressure $p(N)$ that we choose $\it a \; priori$, where $N=\ln a$. 
Let us consider a Lagrangian of the form 
\begin{equation}
 \label{Lagrangian}
 \mathcal{L}(\varphi ,X)= p(\varphi ,X)=f(\varphi)g(h(\varphi)X)-V(\varphi) \; .  
\end{equation}
Note that, with the freedom allowed by the three potentials $f(\varphi)$, $h(\varphi)$ and $V(\varphi)$,
we can independently specify  the equation of state parameter $w$ and the sound speed $c_{\rm s}$ \cite{Bertacca:2008uf}.  In order to reconstruct these potentials we need three dynamical conditions: $a)$ a choice for $p(N)$, $b)$ the continuity equation or, equivalently, the equation of motion (\ref{eq_phi}), $c)$ a choice for $c_s^2(N)$.

Now we look at the four main prescriptions needed to  to obtain the various terms of the Lagrangian (\ref{Lagrangian}).

 \subsection{Equation of state of UDM}\label{EoSUDM}

Assuming $p(N)$ as a function that we choose $\it a \; priori$, Eq. (\ref{coneq}) becomes 
\begin{equation}
\label{gen}
\frac{d\rho(N)}{dN}+3\rho(N)=-3p(N)\;,
\end{equation}
i.e.
\begin{equation}
\label{gen-rho}
\rho(N)=e^{-3N}\left[-3\int^N\left(e^{3N'}p(N')dN'\right) + K\right]\;,
\end{equation}
where $K$ is an integration constant.
In particular, by imposing the condition \\ $\mathcal{L}(X,\varphi)=p(N)$ along 
the classical trajectories on cosmological scales, we obtain \\
$\varphi=\mathcal{L}^{(-1)}(X,p(N))\big|_{\mathcal{M}_{p(N)}}$.
Here, we are constraining the solutions of the equation of motion to live
on a particular manifold $\mathcal{M}_{p(N)}$ [depending on our choice of $p(N)$] embedded 
in the four dimensional space-time.

We assume that the pressure $p$ and energy density $\rho$ of our UDM: 
a) satisfy the null energy condition $\rho + p\geq 0$ \cite{Visser:1997qk, Vikman:2004dc} (here we are not interested in phantom cosmology); 
b) violate the strong condition at late time, i.e. $p<-\rho/3$, in order to reproduce the observed accelerated expansion of the Universe; 
c) imply an attracting fixed point for Eq.\ (\ref{gen}): for $N \to \infty$, $\rho \to \rho_\Lambda$ and $p\to - \rho_\Lambda$, so that $d \rho / d N \to 0$ and $\rho_\Lambda$ plays the role of an unavoidable effective cosmological constant \cite{Ananda:2005xp,Ananda:2006gf,Balbi:2007mz}.

Now,  if $p=-\rho_{\Lambda}$, then $K=(\rho_0-\rho_{\Lambda})$, where $\rho_0$ is the value of $\rho$ today. The freedom provided by
the choice of $K$ is particularly relevant. In fact,
by setting $K=0$, we can remove the term $\rho \propto a^{-3}$ from Eq.\ (\ref{gen-rho}). Alternatively, 
when $K\ne 0$, we always have a term that behaves like pressure-less matter (see Ref.\  \cite{Bertacca:2008uf}), with density today $K=\rho_{\rm m 0}=(\rho_0-\rho_{\Lambda})$.
This simply follows from assuming $p(N)$ as given, so that, in the solution $\rho(N)$ (\ref{gen-rho}), $\rho_{\rm m}(N)= \rho_{\rm m 0} e^{-3N}$ 
represents the matter-like homogeneous solution of (\ref{gen}),
and $p(N)$ generates the particular solution (see \cite{Bertacca:2010ct, Bertacca:2008uf, Bruni-Lazcoz}).

We conclude with the following two remarks:
$i)$  starting from Eqs.~(\ref{gen}) and (\ref{gen-rho})  we can obtain a dark component which can mimic not only a cosmological constant but also any dynamical DE; $ii)$ in order to obtain explicitly the three potentials $f(\varphi)$, $h(\varphi)$ and $V(\varphi)$, $p(N)$ should be chosen in order to have an analytic expression for $\rho(N)$. 

 \subsection{Reconstruction of the potential $f(\varphi)$}\label{Rec-f_phi}

From the Lagrangian $\mathcal{L} = f(\varphi)g(h(\varphi)X)-V(\varphi)$, and Eq.\  (\ref{gen}), we get 
\begin{equation}
\label{gen2}
2X\left[\frac{\partial g(h(\varphi(X,N))X)}{\partial X}\right]f(\varphi(X,N)) = 
p(N)+e^{-3N}\left[-3\int^N\left(e^{3N'}p(N')dN'\right)+K\right]\;.
\end{equation}
The relations (\ref{gen2}) and $\varphi = \mathcal{L}^{(-1)}(X,p(N))\big|_{\mathcal{M}_{p(N)}}$ enable us to determine a connection between the scale factor $a$ and the kinetic term $X$ on the 
manifold $\mathcal{M}_{p(N)}$ and, therefore, a direct mapping between 
the cosmic time and the manifold $\mathcal{M}_{p(N)}$.

Let us define $f(\varphi(X,N))$  in the following way
\begin{equation}
\label{f_XN}
f(\varphi(X,N))=\frac{p(N)+\rho(N)}{2X\left[\partial g(h(\varphi(X,N))X)/\partial X\right]}\Delta(N,X)\;,
\end{equation}
along the classical trajectories on cosmological scales, we deduce that $\Delta(N,X)=1$.
Specifically, let us define $\Delta(N,X)=\widetilde{\Delta}(N,X/\rho(N))$. Then, knowing that $X=(dN/dt)^2(d\varphi/dN)^2/2=\rho(N)(d\varphi/dN)^2/6$, we get 
\begin{eqnarray}
\label{X-N}
X & = & \rho(N)\widetilde{\Delta}^{(-1)}(N)\;, \\
\label{varphi-N}
\varphi  &= & \varphi_i  \pm   \int^N_{N_i} \left(6\widetilde{\Delta}^{(-1)}(N')\right)^{1/2}dN' \;.
\end{eqnarray}
Without any loss of generality, consider the case with the $+$ sign in front of the integral above. Redefining Eqs.\ (\ref{X-N}) and (\ref{varphi-N}) as $X\equiv\mathcal{G}_p(N)$ and $\varphi\equiv\mathcal{Q}_p(N)$, we can write  $f(X,N)=f(\mathcal{G}_p(N),N)= f(\mathcal{G}_p(\mathcal{Q}_p^{(-1)}(\varphi)),\mathcal{Q}_p^{(-1)}(\varphi)) = f(\varphi)$. Obviously, in order to obtain $f(\varphi)$ we must have chosen an  appropriate  $g(h(\varphi)X)$ (see point 3). 

Now, let us assume $\Delta(N,X)$ being of the following form:
\begin{equation}
\Delta(N(a),X)=\frac{\beta  a^3 }{\mu} \frac{\mu \rho(N(a))/(2 \beta \nu) - X}{X}\;,
\end{equation}
where $\beta$ and $\mu$ are two appropriate constants whose choice depends on $p(N)$ and $c_s^2(N)$ and $\nu=\Omega_{\rm m}/\Omega_{\rm DE}$, where $\Omega_{\rm m}=\rho_{\rm m 0}/(3H_0^{2})$ and $\Omega_{\rm DE}=1-\Omega_{\rm m}$. Along the classical trajectories on cosmological scales, we obtain
\begin{eqnarray}
\label{X-N_2}
X(a)&=&\mathcal{G}_p(N(a))=\frac{\mu}{2\nu\beta}\frac{\rho(N(a))}{1+(\mu/\beta)a^{-3}}\;, \\
\label{varphi-N_2}
\varphi(a)&=&\mathcal{Q}_p(N(a))=\left(\frac{4\mu}{3\nu \beta}\right)^{\frac{1}{2}} \sinh^{(-1)}{\left[\left(\frac{\beta}{\mu}\right)^{\frac{1}{2}}a^{\frac{3}{2}}\right]}\;.
\end{eqnarray}
In conclusion, from Eq.\ (\ref{f_XN}), relations (\ref{X-N_2}) and (\ref{varphi-N_2}) allow us to determine explicitly $f(\varphi)$. Obviously if $p=-\rho_\Lambda$ we get the same $X(a)$ and $\varphi(a)$ obtained in Ref.\ \cite{Bertacca:2008uf}.

 \subsection{Reconstruction of the potentials $h(\varphi)$ and $V(\varphi)$}\label{Rec-hV_phi}

In order to get $h(\varphi)$ we have to know $g(\mathcal{X})$, where 
$\mathcal{X}\equiv h(\varphi)X$. From \eqref{csUDMscalar} we have
\begin{equation}\label{cs2_2}
c_s^2(\mathcal{X})=\frac{\partial g(\mathcal{X})/\partial \mathcal{X}}{(\partial g(\mathcal{X})/\partial \mathcal{X})+ 2\mathcal{X}(\partial^2 g(\mathcal{X})/\partial \mathcal{X}^2)} \;, 
\end{equation}
where we must impose $ c_s^2(\mathcal{X}) \ge 0$. In general $c_s^2$ is not bounded from above; however in specific cases  the condition $ c_s^2(\mathcal{X}) \le 1$ must be satisfied. Knowing that $\mathcal{X}=\mathcal{X}(a)$ from Eqs.\   (\ref{X-N_2}) and (\ref{varphi-N_2}), we can choose an appropriate function $c_s^2(\mathcal{X})=\mathcal{R}_p(N(a))$ such that $c_s^2 \ll 1$. Therefore,
\begin{equation}
h(\varphi(a))=\frac{2\nu\beta}{\mu}\frac{1+(\mu/\beta)a^{-3}}{\rho(N(a))} \; \left[c_s^2\right]^{(-1)}(\mathcal{R}_p(N(a)))
\end{equation}
and, through $\mathcal{Q}_p^{(-1)}(\varphi)$, we can reconstruct $h(\varphi)$.   

In order to get $V(\varphi)$ we have to use Eq.\ (\ref{Lagrangian}):
\begin{equation}
V(\varphi(a))=f(\varphi(a))g(\mathcal{X}(a))-p(N(a))\;.
\end{equation} 
Using $\mathcal{Q}_p^{(-1)}(\varphi)$, we can reconstruct $V(\varphi)$.

 \subsection{Simplification of the Lagrangian}\label{Simp-L}

At this point, it is important to stress that the Lagrangian we have obtained may be very complicated. 
In \cite{Bertacca:2008uf} it was shown that there exists a class of Lagrangians having similar kinematical properties: in particular  the equation of state $w$ and speed of sound $c_s$ are invariant under certain transformations within the class. Thanks to this property we are able to remarkably simplify our Lagrangian.

 Specifically, from $\mathcal{L}(X,\varphi)$, we can obtain a new Lagrangian $\mathcal{L}(R(\phi)Y,\phi)$ where
\begin{equation}
Y=\frac{\dot{\phi}^2}{2}=\frac{X}{R(\phi(\varphi))}\quad\quad {\rm and}\quad\quad \phi(\varphi)=\pm
\int^{\varphi}[R(\tilde{\varphi})^{-1/2}d\tilde{\varphi}]+\hat{K} \;,
\end{equation}
with $R(\phi)>0$ and where $\hat{K}$ is an appropriate integration constant. Without any 
loss of generality, consider the case with the $+$ sign in front of the above integral. Defining 
\begin{equation}
\left[R(\phi(\varphi))\right]^{-1/2} = \cosh{\left[ \left(\frac{3\nu\beta}{4\mu}\right)^{1/2}\varphi\right]}\;, 
\end{equation}
we obtain the following relations 
\begin{eqnarray}
\phi(a)&=&\left(\frac{4/3}{\nu a^{-3}}\right)^{1/2}\;,\\
R(\phi(\varphi)) &=& \frac{1}{1+\left[3\nu\beta/(4\mu)\right] \phi^2}\;.
\end{eqnarray}

 \subsection{Prescriptions: further remarks}\label{Comments}

 Let us stress that the receipt described in the previous subsections  is completely general for any $p(N)$ (when $p+\rho \ge 0$) and for any positive $c_s^2(N)$. Obviously, the values of  $c_s^2(N)$ depend strongly on the definition of $g(\mathcal{X})$, through Eq.\ (\ref{cs2_2}).
 
Now additional comments are in order:
\begin{itemize}
\item[i)]  From Section \ref{EoSUDM} we note that, having assumed $\rho_{\rm m}(a)= \rho_{\rm m0}~a^{-3}$  as the pressure-less  solution of (\ref{gen}), we can formally define 
\begin{equation}\label{wDE}
w_{\rm DE}=\frac{p}{\rho-\rho_{\rm m}}\;,
\end{equation}
where $\rho-\rho_{\rm m}$ is the DE-like part on our UDM energy density and $p$ is its pressure.
Interestingly, it is possible to build models with  $w_{\rm DE}<-1$ without violating the null energy condition, see the next Section for explicit example. 

\item[ii)] Choosing an arbitrary equation of state $w(N)$ we obtain 
\begin{eqnarray}
\rho(N)&=&\rho_{i}~e^{-3\int^N(w(N')+1)dN'}\;,\\
p(N)&=&\rho_{i}~w(N)e^{-3\int^N(w(N')+1)dN'}\;,
\end{eqnarray}
where $\rho_i$ is a positive integration constant. Therefore, imposing again the condition $\mathcal{L}(X,\varphi) = 
p[w(N),N]$ along the classical trajectories, i.e.
$\varphi=\mathcal{L}^{(-1)}[X,p(w(N),N)]\big|_{\mathcal{M}_{w(N)}}$, we get
\begin{equation}
\label{gen3}
2X\frac{\partial g(h(\varphi[X,N])X)}{\partial X}f(X,N)=\rho_i~[w(N)+1]
e^{-3\int^N(w(N')+1)dN'}\;.
\end{equation}
Hence, on the classical trajectory we can impose, via $w(N)$, a suitable function $p(N)$ and thus the function $\rho(N)$. Also in this case, from Eq.~(\ref{gen3}) and following 
arguments similar to those described in the points above, we can get the relations
$X\equiv\mathcal{G}_w(N)$, and consequently 
\begin{equation}
\varphi\equiv\mathcal{Q}_w(N)=\pm \int^N\left\{\left[6 \mathcal{G}_w(N')\right]^{1/2}
\left[\rho_{i}~e^{-3\int^{N'}(w(N'')+1)dN''}\right]^{-1/2}dN'\right\} + 
\varphi_i\;.
\end{equation}
With the functions $\mathcal{G}_w(N)$ and $\mathcal{Q}_w(N)$, 
we can write 
$f(X,N)=f(\mathcal{G}_w(N),N)=
f(\mathcal{G}_w(\mathcal{Q}_w^{(-1)}(\varphi)),\mathcal{Q}_w^{(-1)}(\varphi))
= f(\varphi)$.
Then we can find a Lagrangian whose behaviour is determined  
by $w(N)$ and whose speed of sound is determined by the appropriate 
choice of $g(h(\varphi)X)$. 

\item[iii)] Once we have  obtained a Lagrangian, it is important to investigate the kinematic behaviour of the UDM fluid during the radiation-dominated epoch  and, in particular, to choose  appropriate initial conditions in order to ensure that,  during recombination, we obtain solutions of the equation of motion which properly describe $p(N)$ during the subsequent dark epoch. 
\end{itemize}
In the next section, we give an example in which we apply the prescriptions above and consider the explicit solutions in the case of the following Dirac-Born-Infeld type kinetic term:
\begin{equation}
g(h(\varphi)X)=-\sqrt{1-2h(\varphi)X} \;,
\end{equation}
which implies
\begin{equation}
\label{c_s-BI}
c_s^2(h(\varphi)X)=1-2h(\varphi)X\;.
\end{equation}
Moreover, we will choose a suitable $p(N)$ in order to obtain a  class of  UDM models characterized by a fast transition in the equation of state and, in section \ref{sec:perts}, an appropriate $c_s^2(N)$ such that the {\it k}-essence scalar field can cluster. 

\section{A simple UDM scalar field model with fast transition.}\label{p-fast}

We now propose a UDM model with the following equation of state, which we give in parametric form:
\begin{eqnarray}
\label{ptanh}
p(N(a)) & = &  -\frac{\rho_\Lambda}{2} \left\{1 + \tanh\left[\frac{\beta}{3} \left(a^3 - a_{\rm t}^3\right)\right]\right\}\;,\\
\label{tanh}
\rho(N(a)) & = &  \rho_\Lambda \left\{ \frac{1}{2}+\frac{3}{2\beta}a^{-3}\ln\left\{\cosh\left[\frac{\beta}{3} \left(a^3 - a_{\rm t}^3\right)\right]  \right\} + \frac{\rho_{\rm m0}}{\rho_\Lambda} a^{-3}\right\}\;,
\end{eqnarray}
where (\ref{tanh}) follows from  (\ref{ptanh}) by integrating  Eq.\ (\ref{gen-rho}). 
The parameters $a_{\rm t}$ and $\beta$ respectively represent  the scale factor value at which the transition takes place and  the rapidity of the transition. The third parameter of the model is $\rho_\Lambda$:  it is clear from Eq.\ (\ref{tanh}) that in the limit $a \to \infty$ $\rho \to \rho_\Lambda$ and $p\to - \rho_\Lambda $, so that $\rho_\Lambda$ plays the role of  an effective cosmological constant, i.e.\ a fixed point of Eq.\ (\ref{gen}) (see Section \ref{EoSUDM}).  Finally, we note that our UDM model depends on a fourth parameter, $\rho_{\rm m0}/\rho_\Lambda=\Omega_{\rm m}/\Omega_\Lambda$ in Eq.\ (\ref{tanh}) ($\Omega_\Lambda=\rho_\Lambda/(3H_0^2)$).  In comparison with our barotropic fluid model in \cite{Piattella:2009da},  here we have this extra parameter because we have explicitly introduced a matter-like part in the energy density thanks to the integration constant $K$ in (\ref{gen-rho}). 

\begin{figure}[htbp]
\begin{center}
\includegraphics[width=0.496\columnwidth]{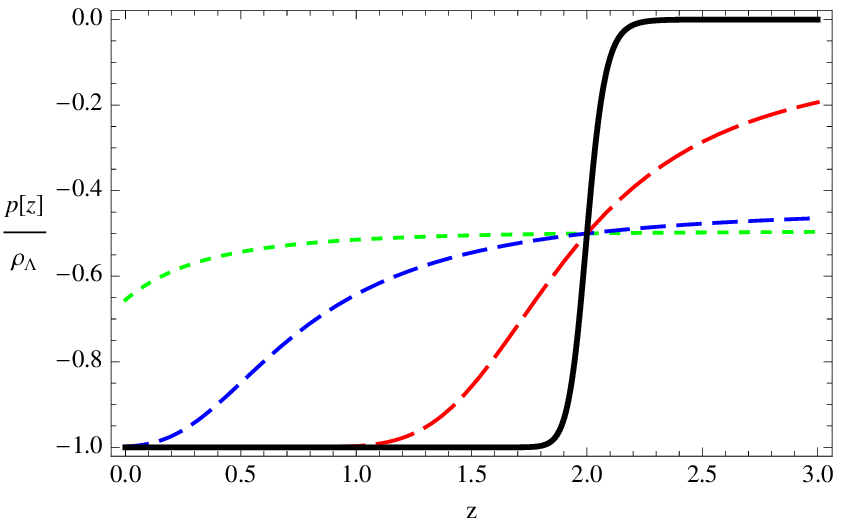}
\includegraphics[width=0.496\columnwidth]{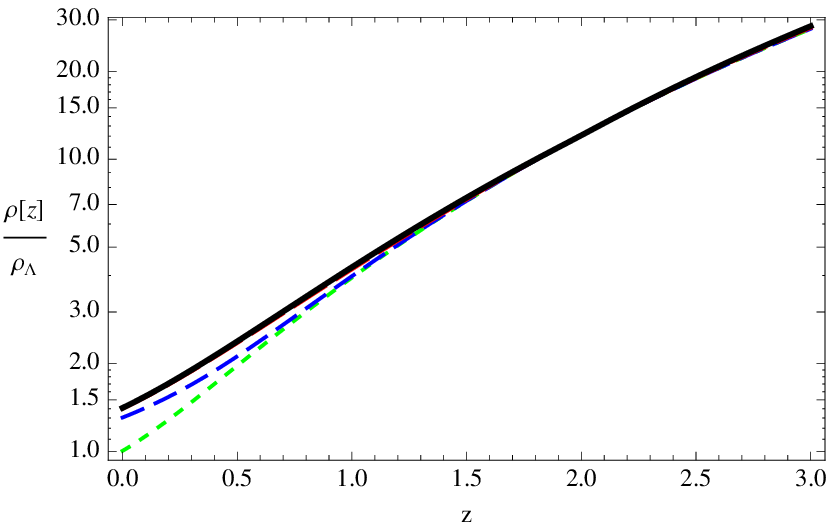}
\includegraphics[width=0.496\columnwidth]{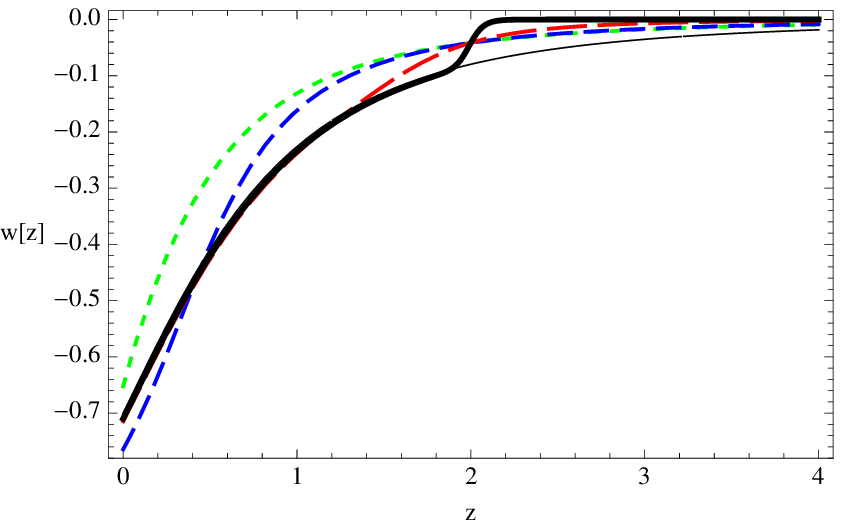}
\caption{Illustrative plot of $p/\rho_\Lambda$,  $\rho/\rho_\Lambda$  and $w$ as functions of the redshift $z$, with $z_{\rm t} = 2$ and  $\Omega_{\rm m}/\Omega_{\Lambda}=3/7$. The lines, from short to long dashes, correspond to $\beta = 1, 10, 100$, respectively; the black solid line corresponds to $\beta = 1000$. For reference, in the bottom panel, we also plot the total $w$ for the $\Lambda$CDM model (thin black line) with $\Omega_{\rm m}/\Omega_{\Lambda}=3/7$ (cf.\  \cite{Ananda:2005xp,Balbi:2007mz, Piattella:2009kt}).}
\label{fig:p-rho-w_z}
\end{center}
\end{figure}

In Fig.\ \ref{fig:p-rho-w_z} we show  the evolution of $p(N(z))$, $\rho(N(a))$ and the equation state parameter $w(N(a))$ as a function of the redshift $z$. An important difference with the barotropic model in \cite{Piattella:2009da} is that the parametric representation (\ref{ptanh})-(\ref{tanh}) is effectively equivalent to an equation of state $p=p(\rho,s)$, where $s$ is
an entropy density. Therefore, in general the condition $p=-\rho$ does not uniquely determine a fixed point of Eq.\  (\ref{gen-rho})  as in the barotropic case. In other words, in  a certain region of parameter space the condition $p=-\rho$ has two solutions: one is the effective cosmological constant $\rho_\Lambda$, the asymptotic value of $\rho$ and $-p$ for $a\to \infty$, the other is a minimum value of energy density  $\rho_*<\rho_\Lambda$ that is attained in the future for a finite value of $a_*>1$, with $p(a_*)=-\rho_*$. For $a>a_*$ the null energy condition is violated, so that the equation of state becomes phantom and $\rho$ grows again, asymptotically approaching $\rho_\Lambda$. In this paper we are interested in mapping the equation of state (\ref{ptanh})-(\ref{tanh}) into a Lagrangian for a scalar field that does not violate the null energy condition, thus we will focus on the relevant region in parameter space (see below).    
\begin{figure}[htbp]
\begin{center}
\includegraphics[width=0.7\columnwidth]{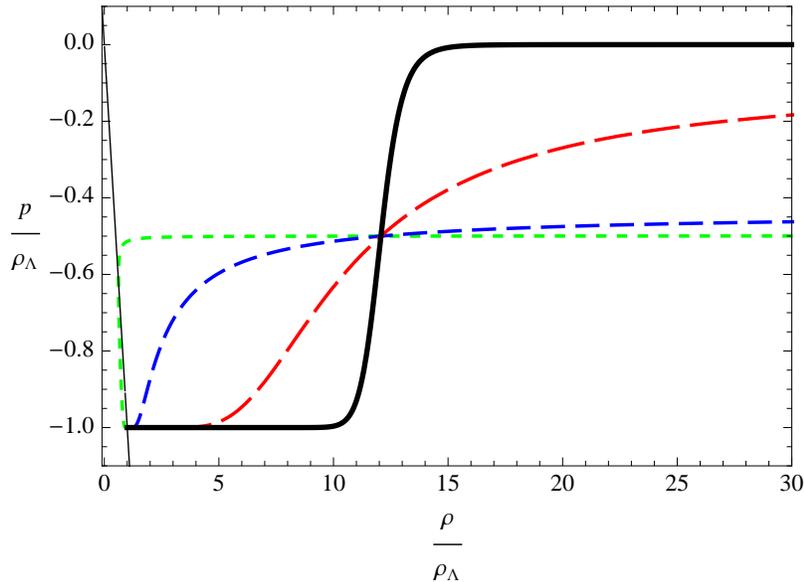}
\caption{Illustrative parametric  plot of $p/\rho_\Lambda$ as a function of $\rho/\rho_\Lambda$, with $z_{\rm t} = 2$ and $\Omega_{\rm m}/\Omega_\Lambda=3/7$. The short to long dashed lines correspond to $\beta = 1, 10, 100$, respectively; the black solid line corresponds to $\beta = 1000$.  For reference we also plot  the $p = -\rho$ line. Note that the latter line is crossed by the $\beta=1$ model (see text). All models asymptotically evolve toward the effective cosmological constant $\rho_\Lambda$.}
\label{fig:prho}
\end{center}
\end{figure}
Fig.\  \ref{fig:prho}  is a parametric plot of $p(a)$  vs.\ $\rho(a)$,  assuming a transition at  $z_{\rm t}=2$  and for  a representative choice of the other parameters. The curve for $\beta=1$ illustrate in particular a case where indeed the parametric equation of state  (\ref{ptanh})-(\ref{tanh}) becomes phantom in the future. 

The parametrization  of Eqs.\   (\ref{ptanh})-(\ref{tanh}) in terms of $\rho_\Lambda$ is mathematically natural, but it can be deduced from Figs.\ \ref{fig:p-rho-w_z}-\ref{fig:prho}  that it is not so practical from a phenomenological point of view; a more useful parametrization is obtained using $\rho_{\rm DE}=\rho_0-\rho_{\rm m0}$, i.e.\ the present value of the DE-like part of our UDM.  
We obtain the following relation:
\begin{equation}\label{ratioLambdaDE}
\frac{\rho_\Lambda}{\rho_{\rm DE}} = \frac{1}{2}+\frac{3}{2\beta}\ln\left\{\cosh\left[\frac{\beta}{3} \left(1 - a_{\rm t}^3\right)\right]  \right\} \;.
\end{equation}
\begin{figure}[htbp]
\begin{center}
\includegraphics[width=0.49\columnwidth]{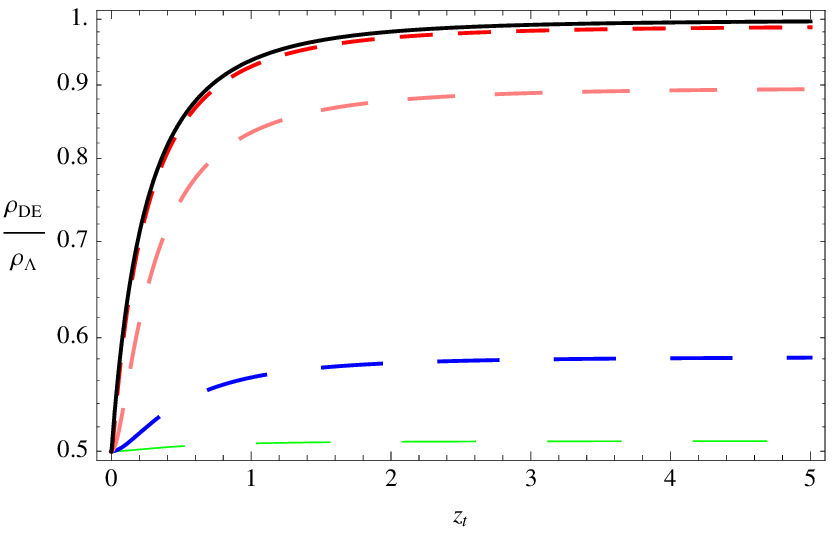}
\includegraphics[width=0.495\columnwidth]{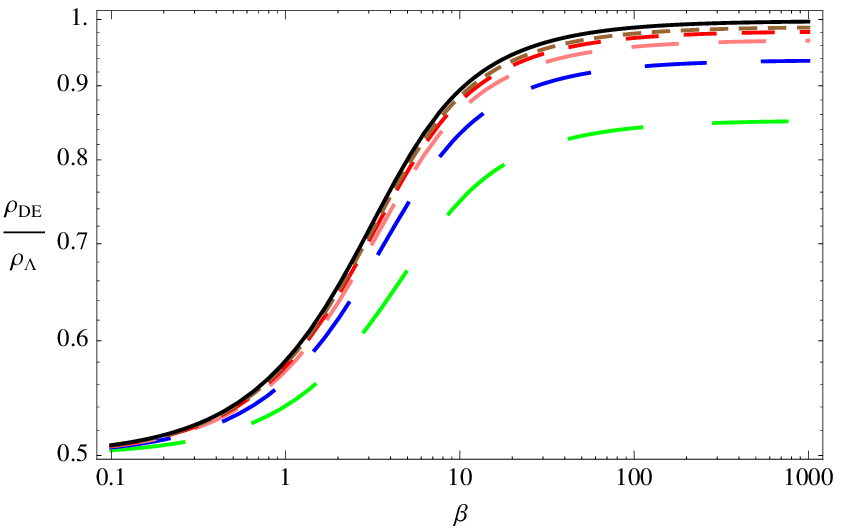}
\caption{Left panel: $\rho_{\rm DE}/\rho_\Lambda$ as function of $z_{\rm t}$, for $\beta = 0.1,1,10,10^2,10^3$  (from bottom to top). Right panel:  $\rho_{\rm DE}/\rho_\Lambda$ as function of $\beta$, for $z_{\rm t} = 0.5, 1, 1.5, 2 , 2.5, 5$ (from bottom to top).}
\label{fig:ratioLambdaDE}
\end{center}
\end{figure}
In Fig.\  \ref{fig:ratioLambdaDE} we plot $\rho_{\rm DE}/\rho_\Lambda$ for different values of $\beta$ and $z_{\rm t}$. Notice that, for $z_{\rm t} >2$ and large $\beta$, we have that $\rho_{\rm DE}/\rho_\Lambda \rightarrow 1$.  

\begin{figure}[htbp]
\begin{center}
\includegraphics[width=0.6\columnwidth]{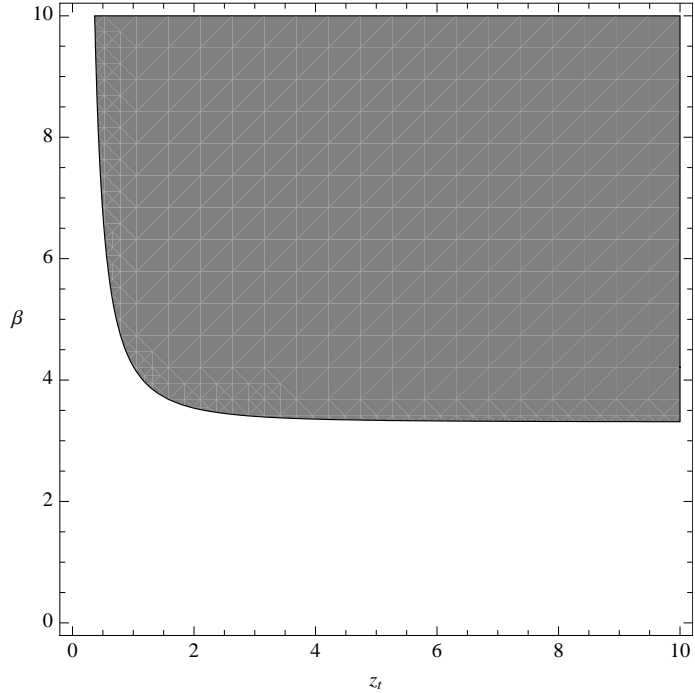}
\caption{The gray region represents the part of the $z_t$-$\beta$ plane where the null energy condition $p+\rho\ge 0$ is satisfied, assuming the representative value $\nu=3/7$.}
\label{fig:at-beta}
\end{center}
\end{figure}

\begin{figure}[h]
\begin{center}
\includegraphics[width=0.496\columnwidth]{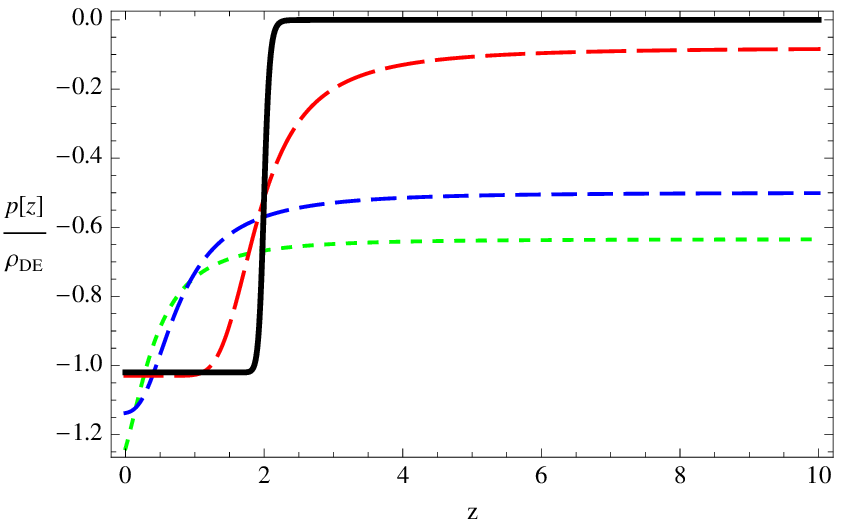}
\includegraphics[width=0.496\columnwidth]{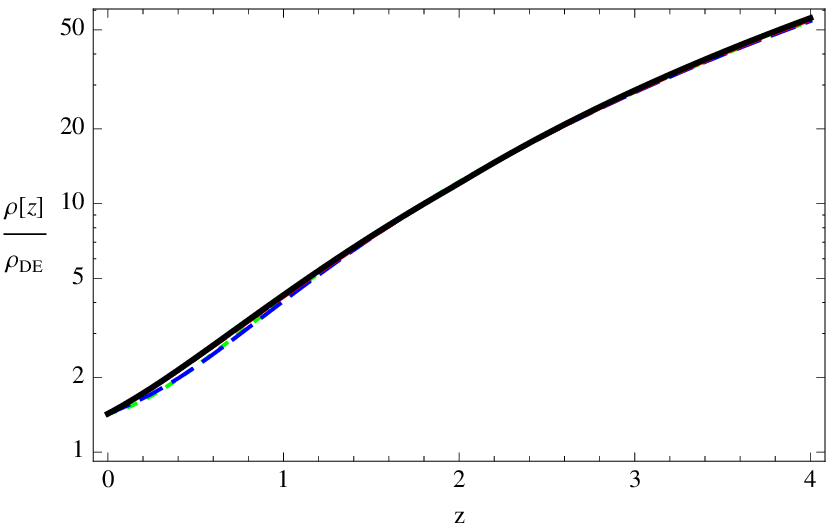} 
\includegraphics[width=0.496\columnwidth]{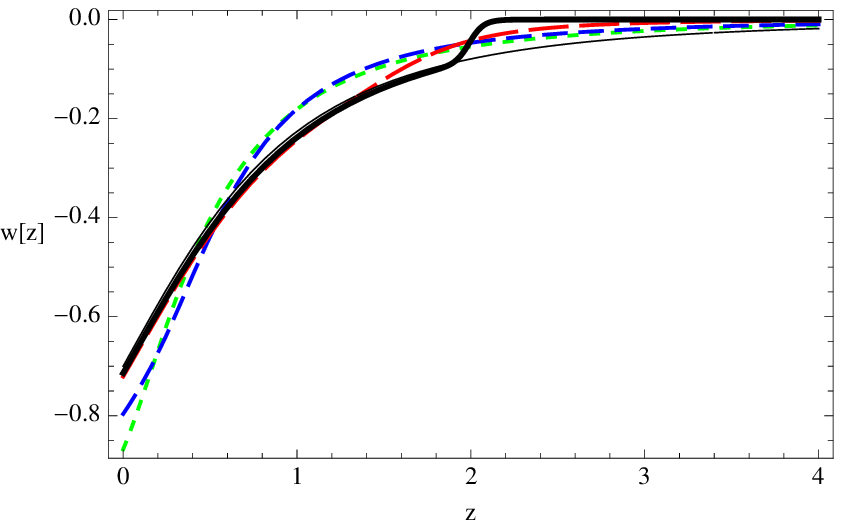}
\caption{Illustrative plot of $p/\rho_{\rm DE}$,  $\rho/\rho_{\rm DE}$  and $w$ as functions of the redshift $z$, with $z_{\rm t} = 2$ and  $\nu=3/7$. The lines, from short to long dashes, correspond to $\beta = 4, 10, 100$, respectively; the black solid line corresponds to $\beta = 1000$. For reference, in the bottom panel, we also plot the total $w$ for the $\Lambda$CDM model (thin black line) with $\Omega_{\rm m}/\Omega_{\Lambda}=3/7$ (cf.\  \cite{Ananda:2005xp,Balbi:2007mz, Piattella:2009kt}).}
\label{fig:prhow2z}
\end{center}
\end{figure}

\begin{figure}[h]
\begin{center}
\includegraphics[width=0.7\columnwidth]{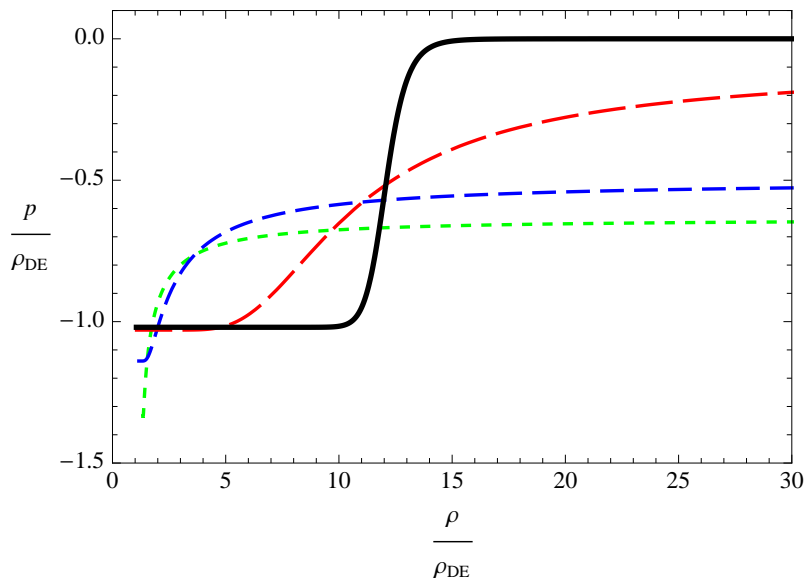}
\caption{Illustrative parametric plot of $p/\rho_{\rm DE}$ as a function of $\rho/\rho_{\rm DE}$, with $z_{\rm t} = 2$ and $\nu=3/7$. The short to long dashed lines  correspond to $\beta = 4, 10, 100$, respectively; the black solid line corresponds to $\beta = 1000$.}
\label{fig:prho2}
\end{center}
\end{figure}

\begin{figure}[h]
\begin{center}
\includegraphics[width=0.496\columnwidth]{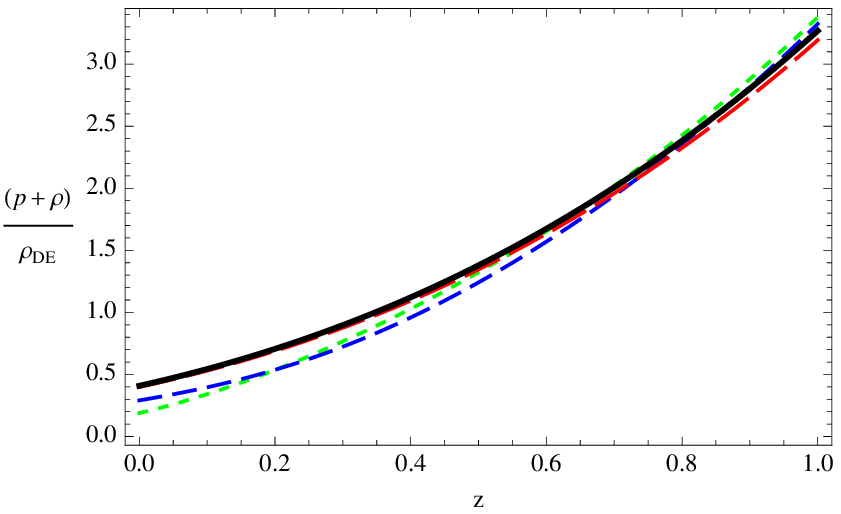}
\includegraphics[width=0.496\columnwidth]{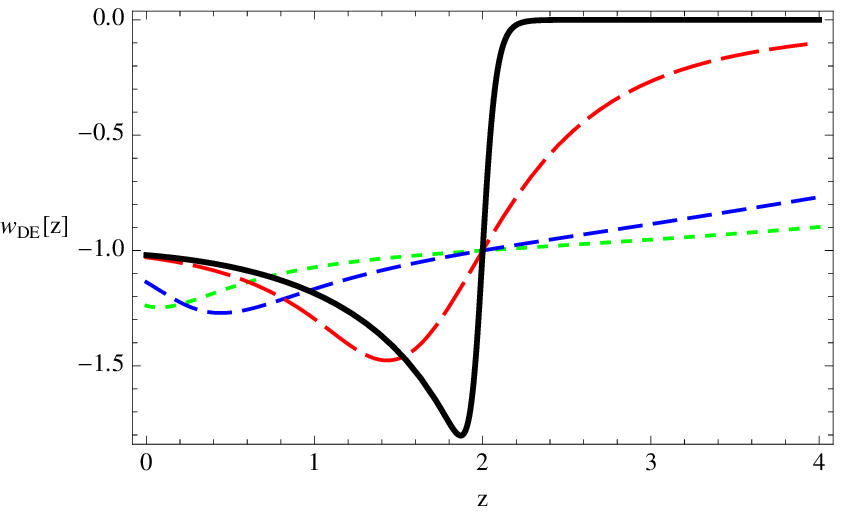}
\caption{Illustrative plots of $(p+\rho)(z)/\rho_{\rm DE}$ and $w_{\rm DE}$ as a function of the redshift $z$, with $z_{\rm t} = 2$ and $\nu=3/7$. In both panels the short to long  dashed lines correspond to $\beta = 4, 10, 100$, respectively; the black solid line corresponds to $\beta = 1000$. Clearly $(p+\rho)>0$ always, i.e. the null energy condition is never violated.}
\label{fig:p+rho2z}
\end{center}
\end{figure}
\begin{figure}[h]
\begin{center}
\includegraphics[width=0.49\columnwidth]{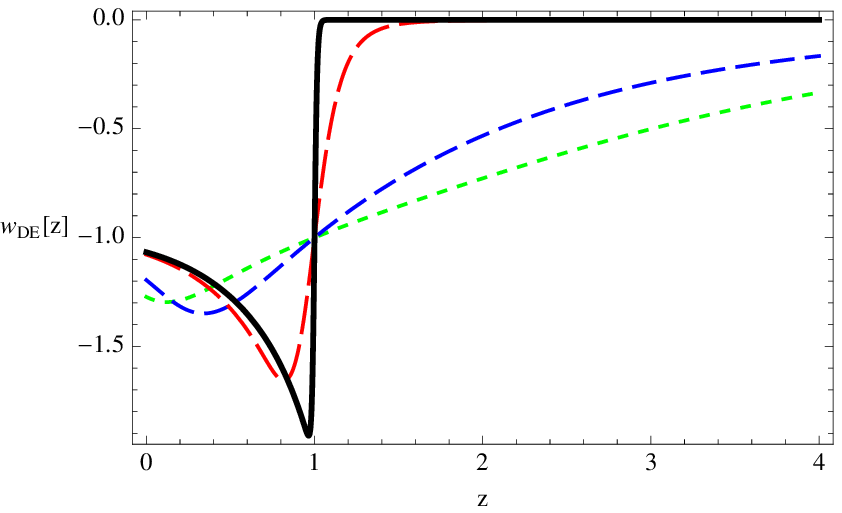}
\includegraphics[width=0.5\columnwidth]{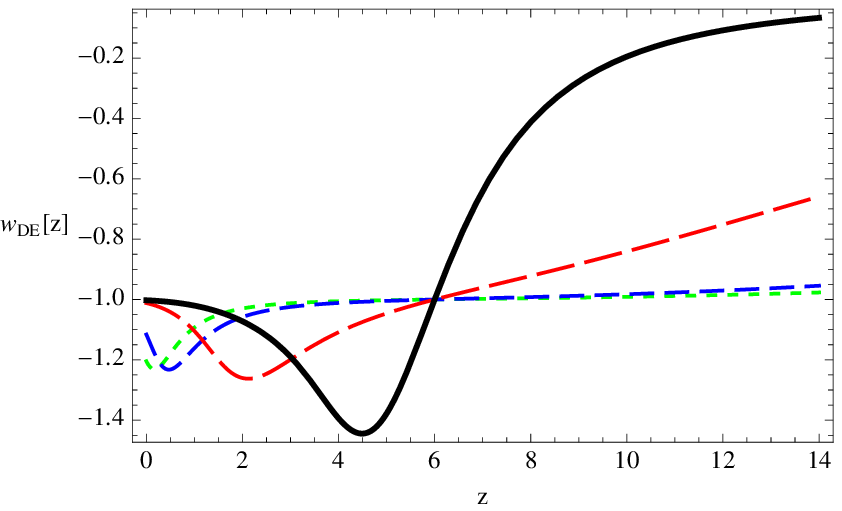}
\caption{$w_{\rm DE}$ as a function of the redshift $z$, with  $\nu=3/7$ and $z_{\rm t} = 1$ (left panel) and $z_{\rm t} = 6$ (right panel). In both panels the short to long dashed lines correspond to $\beta = 5, 10, 100$, respectively; the black solid line corresponds to $\beta = 1000$.  Without violating the null energy condition our models produce $w_{\rm DE}$ after the transition.}
\label{fig:wDE1-6z}
\end{center}
\end{figure}

Now, combining Eq.\  (\ref{ratioLambdaDE}) with Eqs.~\eqref{ptanh}-\eqref{tanh}, $p(a)$ and $\rho(a)$ can be re-expressed as
\begin{eqnarray}\label{p2}
p(a)&=&  -\rho_{\rm DE} \frac{\left\{1 + \tanh\left[(\beta/3) \left(a^3 - a_{\rm t}^3\right)\right]\right\}}{1+ (3/\beta)  \ln\left\{\cosh\left[(\beta/3) \left(1 - a_{\rm t}^3\right)\right]  \right\}}\;,  \\ \label{rho2}
\rho(a)&=&\rho_{\rm DE}\left\{\frac{ 1+(3/\beta)a^{-3}\ln\left\{\cosh\left[(\beta/3) \left(a^3 - a_{\rm t}^3\right)\right] \right\}}{1+ (3/\beta)  \ln\left\{\cosh\left[(\beta/3) \left(1 - a_{\rm t}^3\right)\right]  \right\}}+\nu a^{-3}\right\}\;,
\end{eqnarray}
where $\nu = \rho_{\rm m0}/\rho_{\rm DE}=\Omega_{\rm m}/\Omega_{\rm DE}$ (see  Section \ref{Rec-f_phi}). A good guess for the value of 
this parameter is 
$\nu=3/7$:  assuming this value, in Fig.\ \ref{fig:at-beta} we plot the region of the $z_t-\beta$ plane where the null energy condition $p+\rho\ge 0$ is satisfied. 

In Fig.\  \ref{fig:prhow2z}, we plot $p(z)$, $\rho(z)$, $w(z)$, as functions of the redshift $z$ for different values of $\beta$ and $z_{\rm t}$ and for $\nu=3/7$. 

Finally, in Fig.\ \ref{fig:prho2} we show  a parametric plot of $p(a)$  vs.\ $\rho(a)$,  assuming a transition at  $z_{\rm t}=2$  and for $\nu=3/7$. From Figs.~\ref{fig:p+rho2z}  and \ref{fig:wDE1-6z} and Eq.\ (\ref{wDE}) we see  that for $z <  z_{\rm t}$ we can obtain $w_{\rm DE}  < -1$ without violating the null energy condition [see comment ${\rm i)}$ in Section \ref{Comments}]. 
When $\beta > 10^3$ and $z_{\rm t} > 2$, we have that $\rho_{\rm DE} \simeq \rho_\Lambda$; in this case, our class of UDM models is characterised by a fast transition regime and today  $w_{\rm DE} \simeq -1$.
However, this is not the case for $\beta < 10^3$ or $z_{\rm t} < 2$, so that $w_{\rm DE}$ can be significantly smaller than $-1$ even at small redshift.

\section{Analysis of the Jeans wave number and the gravitational potential}\label{sec:perts}

Knowing $p(N)$  and $\rho(N)$,  following the prescriptions of section \ref{prescription} we still need to choose a suitable $c_s^2$ in order to  obtain the Lagrangian $\mathcal{L}(Y,\phi)$ that will completely specify our UDM scalar field model. This Lagrangian will  reproduce our choice of $p(N)$, $\rho(N)$ and $c_s^2(N)$ on-shell, i.e. along the classical trajectories on cosmological scales (see Appendix \ref{A}).
In this section we choose a suitable $c_s^2$ and, via the equation state (\ref{p2})-(\ref{rho2}), we study the evolution of the Jeans wave-number and the gravitational potential. 

\subsection{Speed of sound and Jeans scale}

We assume a speed of sound of the following form:
\begin{equation}\label{cs}
 c_s^2:= \frac{c_\infty^2}{1+\left(\nu a^{-3}/c_\infty^2 \right)^{n}}\;,
\end{equation}
where $c_\infty > 0$ and $n \ge 0$  are free parameters, the former representing the asymptotic future speed of sound. 

\begin{figure}[htbp]
\begin{center}
\includegraphics[width=0.496\columnwidth]{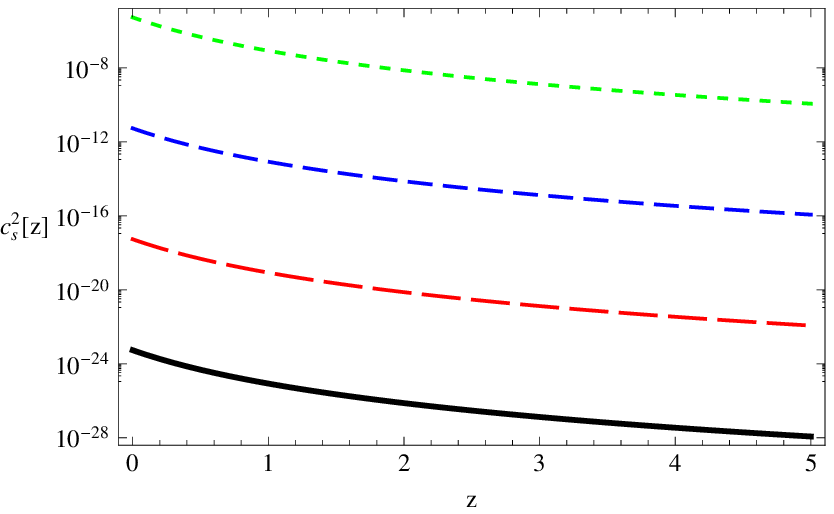}
\includegraphics[width=0.496\columnwidth]{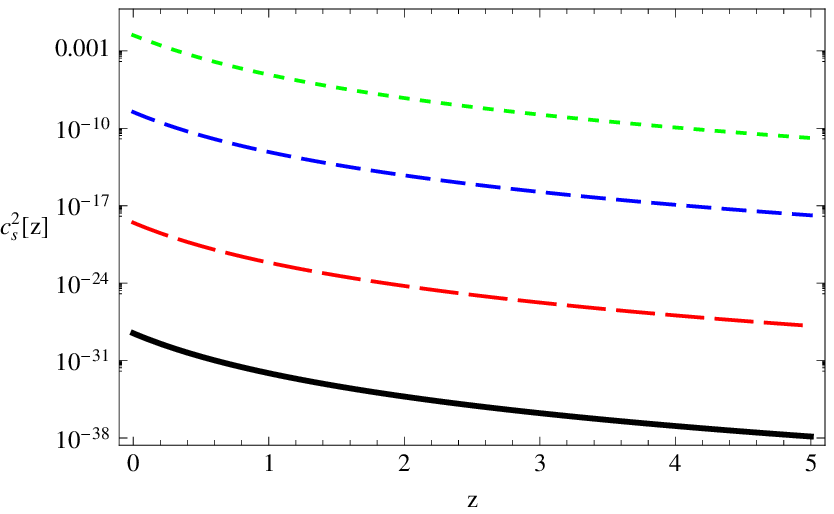}
\caption{Illustrative plot of $c_s^2$ as a function of the redshift $z$, with $\nu=3/7$ and $n = 2$ (left panel) and $n = 4$ (right panel). 
Left panel: the short to long dashed lines respectively correspond to  $c_\infty = 10^{-1}, 10^{-2}, 10^{-3}$ and the black solid line corresponds to $c_\infty = 10^{-4}$. 
Right panel: the short to long dashed lines  respectively correspond to $c_\infty = 0.5, 10^{-1}, 10^{-2} $ and the black solid line corresponds to $c_\infty =  10^{-3}$.
}
\label{fig:cs-n2_4-cinfty}
\end{center}
\end{figure}

\begin{figure}[htbp]
\begin{center}
\includegraphics[width=0.7\columnwidth]{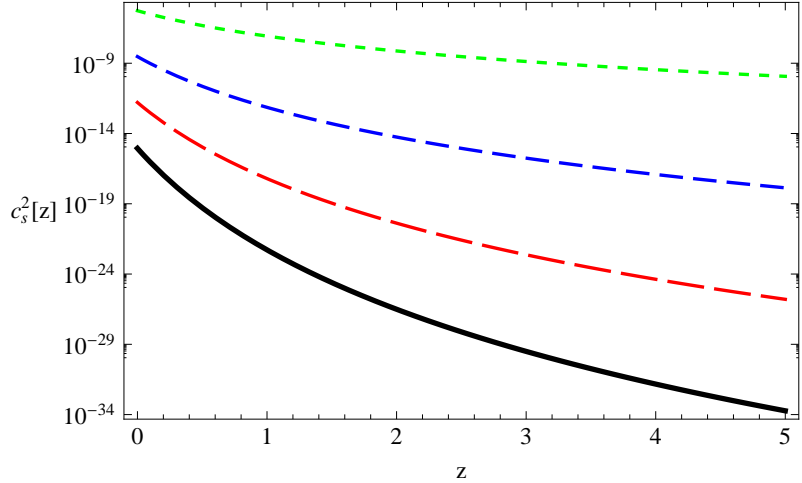}
\caption{Illustrative plot of $c_s^2$ as a function of the redshift $z$, with $c_\infty  = 0.1$ and $\nu=3/7$. The lines, from short to long dashes, respectively correspond to $n = 2, 4, 6$ and the black solid line corresponds to $n = 8$.}
\label{fig:cs-cinfty01-n}
\end{center}
\end{figure}

In Figs.\ \ref{fig:cs-n2_4-cinfty} and \ref{fig:cs-cinfty01-n} we have plotted $c_{\rm s}^2$ for different values of $c_\infty $ and $n$. We can deduce that, for practicality and in order to reduce the number of free parameters, we can choose $n=4,5$ and $c_\infty=0.1$.  In this case we obtain a value for the speed of sound that is suitably small at all times, without requiring any fine tuning.

Let us now analyse the Jeans scale; rather then directly considering the physical Jeans length $\lambda_{\rm J}$ we look at the Jeans wave number $ k_{\rm J} = 2 \pi a/\lambda_{\rm J}$. 
Starting from Eq.\ (\ref{diff-eq_u}), the squared Jeans wave-number is defined as follows \cite{Bertacca:2007cv}:
\begin{equation}\label{kJ2}
k^{2}_{\rm J} := \left|\frac{\theta''}{c_{\rm s}^{2}\theta}\right|\;.
\end{equation}
This is a crucial quantity in determining the viability of a UDM model, because of its effect on perturbations, which is then revealed in observables such as the CMB and matter power spectra. 
Indeed, any UDM model should satisfy the condition  $k_{\rm J}^2 \gg k^2$ for all the scales of cosmological interest, in turn giving an evolution for the gravitational potential $\Phi(\eta,{\bf x}) \propto  \Phi(\eta,k) \exp\left(i{\bf k} \cdot {\bf x}\right)$ of the following type:
\begin{equation}\label{kjggksol}
 \Phi(\eta,k) \simeq A_{\rm k}\left[1 - \frac{H(\eta)}{a(\eta)}\int a(\hat{\eta})^2 d\hat{\eta}\right]\;,
\end{equation}
where $A_{\rm k} = \Phi\left(0,k\right)T_{\rm m}\left(k\right)$, $\Phi\left(0,k\right)$ is the primordial gravitational potential at large scales, set during inflation,  and $T_{\rm m}\left(k\right)$ is the present time matter transfer function, see e.g.\ \cite{Dodelson:2003ft}.

From (\ref{kJ2}), the explicit form of the Jeans wave number for  a scalar field UDM model turns out to be 
\begin{equation}\label{kJ2analytic}
 k_{\rm J}^{2} = \frac{3}{2}\frac{\rho}{(1 + z)^2}\frac{(1 + w)}{c_{\rm s}^2}\left|\frac{1}{2}(c_{\rm ad}^2 - w) - \rho\frac{(c_{\rm ad}^2)'}{\rho'} + \frac{3(c_{\rm ad}^2 - w)^2 - 2(c_{\rm ad}^2 - w)}{6(1 + w)} + \frac{1}{3}\right|\;,
\end{equation}
where $c_{\rm ad}^2=p'/\rho'$. Comparing with the Jeans wave number for a barotropic fluid (see Eq.\ (3.3) in \cite{Piattella:2009kt}), for which $c_{\rm s}^2=c_{\rm ad}^2$, the only difference is  the overall $1/{c_{\rm s}^2}$ factor replacing $1/{c_{\rm ad}^2}$: this gives extra freedom in building a suitable  $ k_{\rm J} $.

\begin{figure}[h]
\begin{center}
\includegraphics[width=0.496\columnwidth]{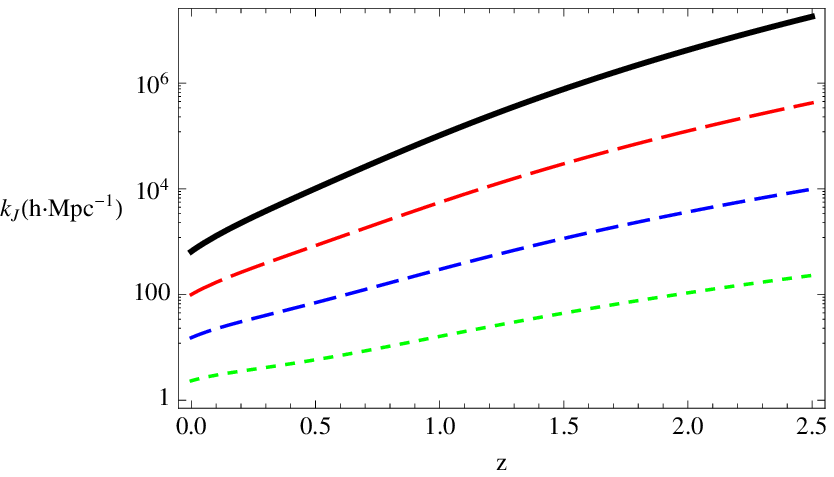}
\includegraphics[width=0.496\columnwidth]{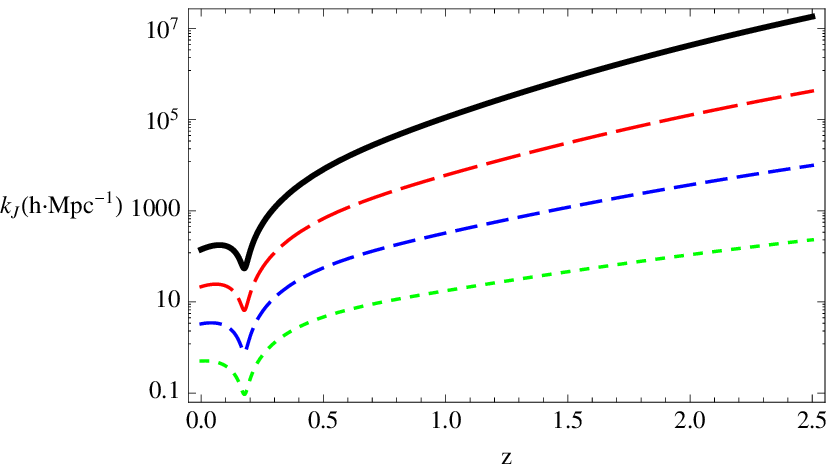}
\includegraphics[width=0.496\columnwidth]{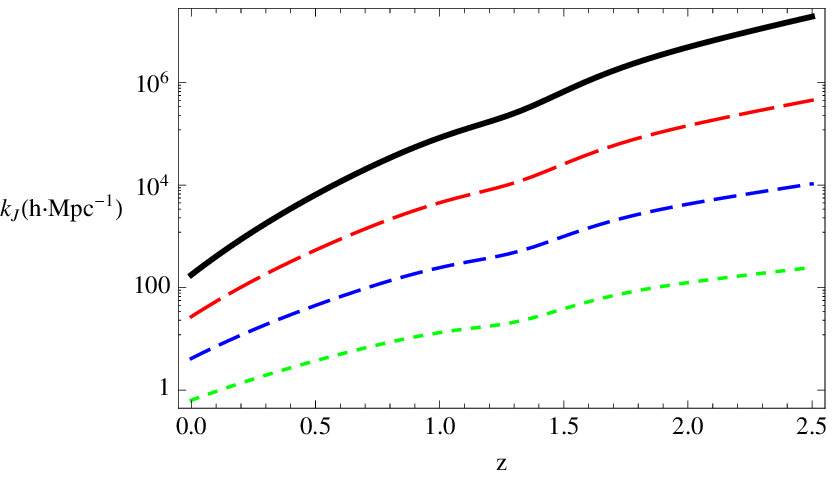}
\includegraphics[width=0.496\columnwidth]{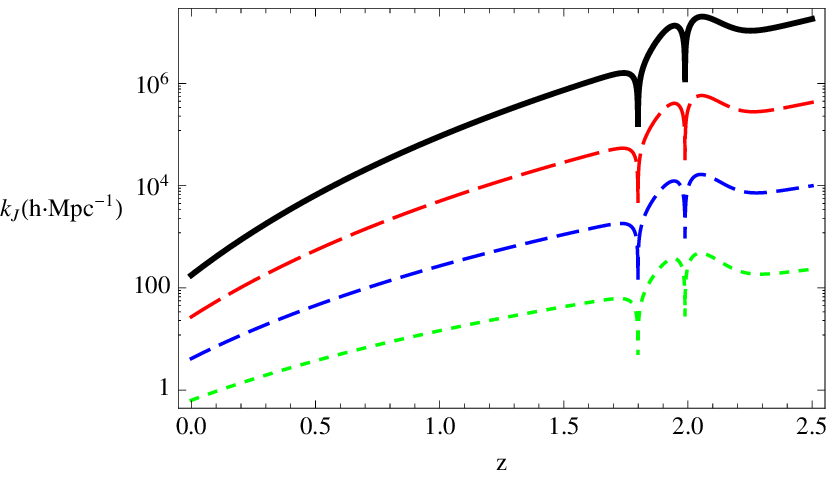}
\caption{The Jeans wave-number $k_{\rm J}$ (in $h$ Mpc$^{-1}$ units)  assuming $\nu=3/7$, $z_{\rm t} = 2 $,  $c_\infty  = 0.1$ and $\beta = 4$ (left-top panel), $\beta = 12.1$ (right-top panel), $\beta = 100$ (left-bottom panel) and $\beta = 1000$ (right-bottom panel). The lines, from short to long dashes, respectively correspond to $n = 3, 4, 5$ and the black solid line corresponds to $n = 6$.}
\label{fig:kj}
\end{center}
\end{figure}

Clearly, we can indeed obtain a large $k_{\rm J}^{2}$  when $c_{\rm s}^2 \to 0$; in addition, when $c_{\rm ad}^2$ changes rapidly around $z_{\rm t}$, i.e.\ when the above expression is dominated by the $\left[\rho\; (c_{\rm ad}^2)'/\rho' \right]$ term, we can have a fast trasition. Therefore, we can conclude that, for $\beta < 1000$ and before and after the fast transition for $\beta > 1000$, it is crucial that $c_{\rm s}^2$ be sufficiently small. 
Defining  $c_{\rm s}^2$ as in Eq.\ (\ref{cs}), in Figs. \ref{fig:kj} we plot the Jeans wave-number for the  representative case $z_t=2$ and $c_\infty=0.1$, for various values of $\beta$ and $n$. 

From the right panels of Fig.\  \ref{fig:kj}, we can note that for $\beta=12.1$ and $1000$ the Jeans wave number $k_{\rm J}$ momentarily vanishes. In general, around these points the corresponding Jeans length becomes very large, possibly causing all sort of problems to perturbations, with effects on structure formation in the UDM model. On the other hand, for sufficiently small $c_{\rm s}$ we note that  {\it i)} in general the Jeans wave number becomes larger and {\it ii)} it becomes vanishingly small for extremely short times, so that the effects caused by its  vanishing  are sufficiently negligible,  as we are going to show in the next subsection when we will analyze the gravitational potential $\Phi$.
Therefore, in building a phenomenological model, we can choose its parameter values  in order to always satisfy the condition  $k \ll k_{\rm J}$ for all $k$ of cosmological interests to which linear theory applies. In other words, we can always build our model in such a way that  all scales smaller than the Jeans length $\lambda \ll \lambda_{\rm J}$ correspond to those in the non-linear regime, i.e.\ scales beyond the range of applicability of the model.
So, for these scales, no conclusions on the behaviour of perturbations can be derived from the linear theory.  Indeed, to investigate these scales, one needs to go beyond the perturbative regime investigated here, possibly also increasing the sophistication  of the UDM model in order to properly take into account  the greater complexity of small scale non-linear physics.
Moreover, from Fig.\ \ref{fig:kj} we can finally conclude that, for $n=4$ or $5$ and $c_\infty=0.1$, we obtain a acceptable Jeans wave number at all times.

\subsection{The gravitational potential}\label{sec:grav-pot}

From Eq.\ (\ref{diff-eq_u}) let us write explicitly the differential equation for the gravitational potential $\Phi$:
\begin{equation}
\label{diff-eq_Phi}
\frac{d^2 \Phi(a,k)}{d a^2}+\left(\frac{1}{\mathcal{H}} \frac{d \mathcal{H}}{d a} + \frac{4}{a}+ 3 \frac{c_{\rm ad}^2}{a}\right)\frac{d \Phi(a,k)}{d a}+\left[ \frac{2}{a \mathcal{H}}\frac{d \mathcal{H}}{d a} +  \frac{1}{a^2}(1+3 c_{\rm ad}^2)+\frac{c_s^2 k^2}{a^2 \mathcal{H}^2} \right] \Phi (a,k) =0\;,
\end{equation}
where $\mathcal{H}=a^2H^2$ and plane-wave perturbation, $\Phi(a,{\bf x}) \propto  \Phi(a,k) \exp\left(i{\bf k} \cdot {\bf x}\right)$ have been assumed.

\begin{figure}[h]
\begin{center}
\includegraphics[width=0.49\columnwidth]{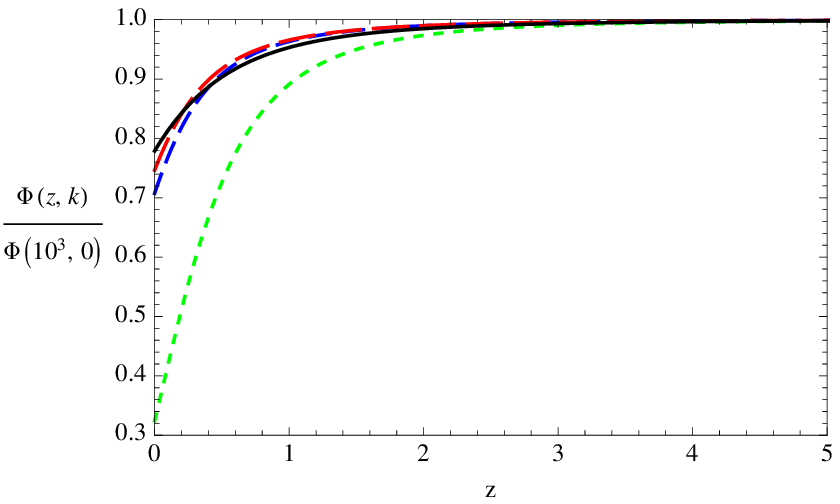}
\includegraphics[width=0.496\columnwidth]{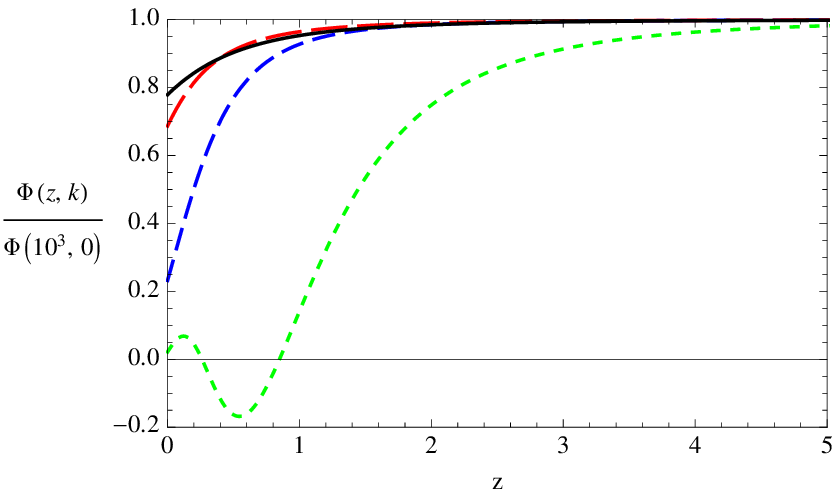}
\caption{Gravitational potential $\Phi(z, k)$ as a function of the redshift $z$, assuming $\nu= 3/7$, $z_{\rm t} = 2 $,  $c_\infty  = 0.1$ and $\beta = 4$. The lines, from short to long dashes, respectively correspond to $n = 1, 1.5, 2$. For comparison, the black solid line represents $\Phi$ the $\Lambda$CDM model  with $\Omega_{\rm m}/\Omega_\Lambda = 3/7$. Left panel: $k=0.05$ $h$ Mpc$^{-1}$; right panel: $k=0.2$ $h$ Mpc$^{-1}$.}
\label{fig:Phi-k}
\end{center}
\end{figure}

In Figs.\  \ref{fig:Phi-k} we plot the gravitational potential for different values of $n$, $\beta=0.1$, $z_{\rm t} = 2 $,  $c_\infty  = 0.1$ and for $k= 0.05$  $h$ Mpc$^{-1}$ and $k= 0.2$  $h$ Mpc$^{-1}$ .  We note that, for $k \sim k_{\rm J}$,   $\Phi(a,k)$ starts to oscillate and decays, thus preventing structure formation. On the other hand for $n>2$ and for small values of $\beta$, we obtain a shape of the gravitational potential very close to that of the $\Lambda$CDM model.

\begin{figure}[h]
\begin{center}
\includegraphics[width=0.495\columnwidth]{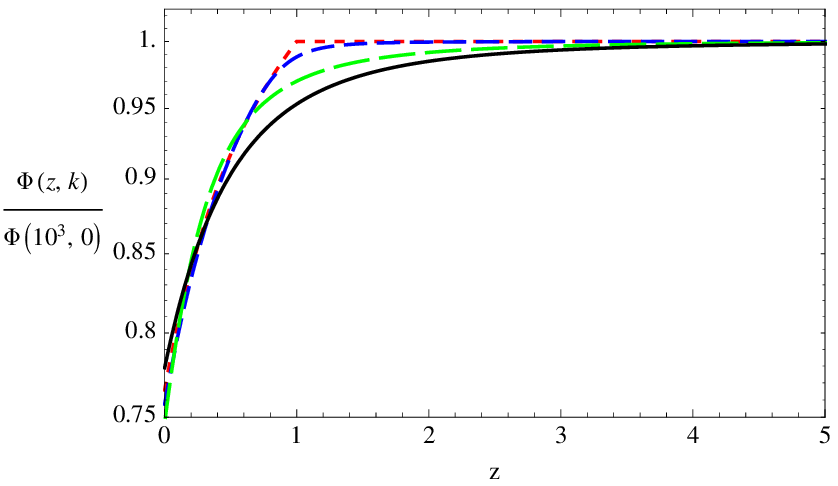}
\includegraphics[width=0.495\columnwidth]{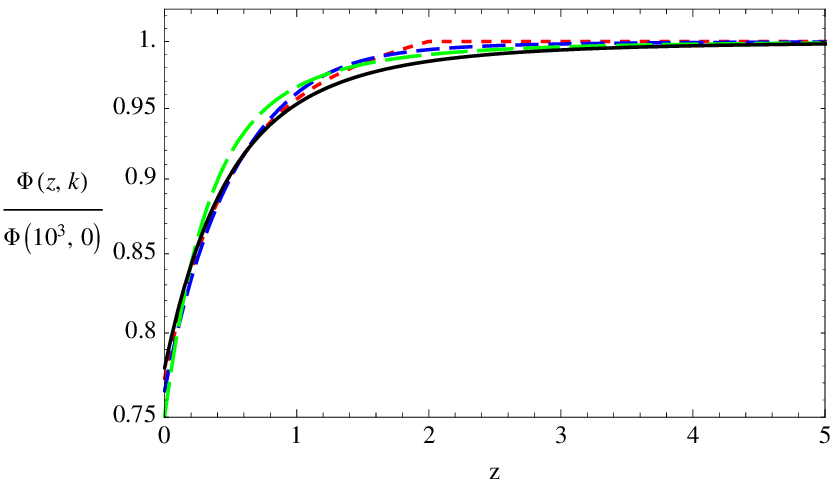}
\includegraphics[width=0.495\columnwidth]{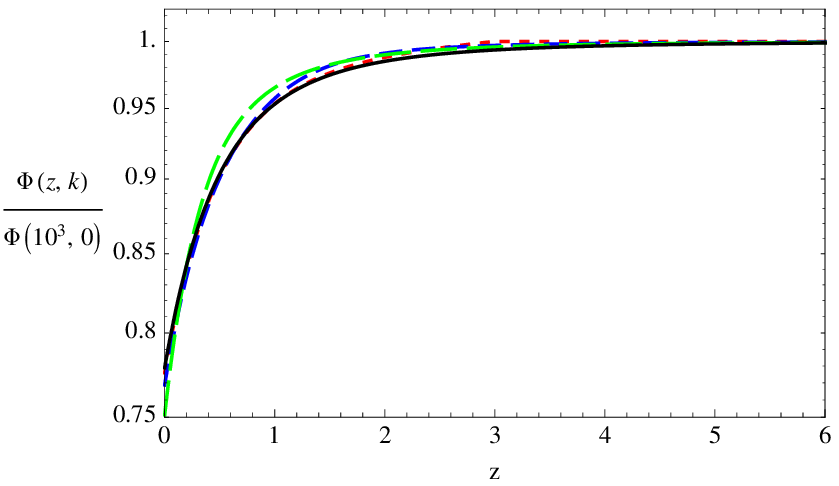}
\includegraphics[width=0.495\columnwidth]{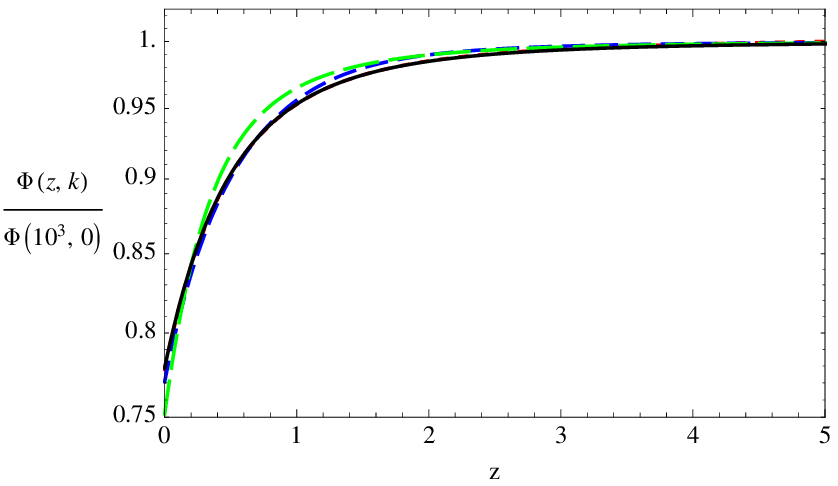}
\caption{Illustrative plots of the gravitational potential $\Phi(z,k)$ as a function of the redshift $z$, for $k=0.2$ $h$ Mpc$^{-1}$, $\nu=3/7$, $z_{\rm t} = 1 $,  $c_\infty  = 0.1$ and $n = 4$. For comparison, the black solid line represents $\Phi$ in the $\Lambda$CDM model with $\Omega_{\rm m}/\Omega_\Lambda = 3/7$. 
Left-top panel: $z_{\rm t} = 1 $. The lines, from short to long dashes, respectively correspond to $\beta = 10^4, 50, 5$.
Right-top panel:  $z_{\rm t} = 2$. The lines, from short to long dashes, respectively correspond to $\beta = 10^4, 50, 4$.
Left-bottom panel:  $z_{\rm t} = 3 $.  The lines, from short to long dashes, respectively correspond to $\beta = 10^4, 50, 4$.
Right-bottom panel: $z_{\rm t} = 5 $. The lines, from short to long dashes, respectively correspond to $\beta = 10^4, 50, 4$.}
\label{fig:Phi-beta}
\end{center}
\end{figure}

Now, assuming for  simplicity $n=4$ and $c_\infty=0.1$ (in order to have an acceptable $k_{\rm J}$) we analyse the  gravitational potential and investigate how it depends on the background parameters $\beta$ and $z_{\rm t}$ (or, equivalently, $a_{\rm t}$). As we know from Fig.\  \ref{fig:prhow2z}, for $\beta < 100$ the value of $z_{\rm t}$ practically loses the meaning of scale parameter at the transition. For these range of parameters, we can also observe this effect in Fig.\ \ref{fig:Phi-beta}, where we have plotted $\Phi(z,k)$ for different values of $\beta$ and $z_{\rm t}$.

Moreover, from these panels we can note another interesting effect. For large values of $\beta$ and for small values of $z_{\rm t}$,  $\Phi(z,k)$ is constant in time (as it should be in a pure matter Einstein De Sitter model) and assumes the same value of $\Phi(0,k)$ until $z\sim z_{\rm t}$ and then for $z < z_{\rm t}$ its value quickly goes down, eventually intersecting the potential  of  the $\Lambda$CDM model. This happens because the background starts to ``feel'' the effective cosmological constant only for $z < z_{\rm t}$. It is important to stress that this evolution of $\Phi$ could produce a strong Integrated Sachs Wolfe (ISW) effect: however, this would only be due to the particular evolution and would not depend from $c_{\rm s}^2$.  Obviously, this effect is weaker if we set $z_{\rm t} > 2$ and completely negligible for $z_{\rm t} \ge 5$, i.e.\ these UDM models become indistinguishable from the $\Lambda$CDM model, cf.\ \cite{Piattella:2009kt}.

In conclusion, the suitability of models with $4 \lesssim  \beta<1000$, as well as that of models with $\beta>1000$ and $z_{\rm t} < 2$,  
need a study of the matter and CMB power spectra, which will reserve for the future. On the other hand, for $\beta>1000$ and assuming $z_{\rm t}>2$, i.e.\  for an early enough fast transition, the above analysis shows that our UDM models should be compatible with observations.

In order to compare the predictions of our UDM model with observational data, we have to  define the UDM density contrast as $\delta := \delta\rho/\rho_{\rm A}$ \cite{Pietrobon:2008js}, where here $\rho_{\rm A}=\rho-\rho_{\Lambda}$ is the clustering ``aether'' part of the UDM component \cite {Ananda:2005xp,Linder:2008ya}. Indeed, our equation of state admits an asymptotic ($a \rightarrow \infty$)   effective ``cosmological constant" \cite{Ananda:2005xp}  which we have already defined as $\rho_\Lambda$ in section \ref{p-fast}.  In this case, starting from the perturbation theory that we outlined in section \ref{sec:bgperteq}, we can infer the link between the density contrast and the gravitational potential via the Poisson equation for scales smaller than the cosmological horizon and $z < z_{\rm rec}$, where $z_{\rm rec}$ is the recombination redshift ($z_{\rm rec} \approx 10^{3}$) in the following way:
\begin{equation}
\label{delta}
 \delta\left(k,z\right)  = -\frac{2 k^{2}\Phi(z,k)\left(1 + z\right)^2}{\rho_{\rm A}}\;.
\end{equation}

\section{Conclusions}\label{sec:concl}


The last decade of observations of large scale structure \cite{Allen:2004cd,Tegmark:2006az,Percival:2006kh,Percival:2007yw,Percival:2009xn,Reid:2009xm}, the search for type
Ia supernovae (SNIa) \cite{Riess:1998cb,Perlmutter:1998np,Riess:1998dv, Amanullah:2010vv} and the measurements of the CMB anisotropies
\cite{Spergel:2003cb,Larson:2010gs, Komatsu:2010fb} are very well explained by assuming  that two dark
components govern the dynamics of the Universe. They are DM, thought to be
the main responsible for structure formation, and an additional DE component
that is supposed to drive the measured cosmic acceleration \cite{ Tsujikawa:2010sc, Amendola:1272934, Copeland:2006wr,Peebles:2002gy,Padmanabhan:2002ji}.
However, it should be recognised that, while some form of CDM is independently expected to exist within any modification of the Standard Model of high energy physics, the really compelling reason to postulate DE has been the acceleration in the cosmic expansion.
It is mainly for this reason that it is worth investigating the hypothesis that CDM and DE are two aspects of a single UDM component,  see e.g.\  \cite{Bertacca:2010ct, Amendola:1272934}.

In this paper  we have developed and generalized the technique to construct scalar field UDM models  proposed in Ref.\ \cite{Bertacca:2008uf} and we have focused on  Lagrangians with non-canonical kinetic term to obtain models where a single component can  mimic the dynamical effects of Dark Matter and Dark Energy and, at the same time, has a sound speed small enough to allow for structure formation.

In the second part  of the paper, we have built UDM models which can produce a fast transition,  similarly to the barotropic fluid  UDM models  we proposed in \cite{Piattella:2009kt}. However, while the background evolution can be very similar in the two cases,  the  perturbations  are naturally adiabatic in fluid models, while in  the scalar field  case they are necessarily  non-adiabatic \cite{Bardeen:1983qw,Bruni:1991kb, Mukhanov:2005sc}, cf.\ \cite{Hu:1998kj, Bardeen:1980kt, Kodama:1985bj}.
This new scalar field model allows to escape the problem of the fine-tuning on the parameters which usually appeared in many previous UDM Lagrangians.

First of all, an interesting feature of our models is that, for $z <  z_{\rm t}$, $w_{\rm DE}$ can be $<-1$ without violating the null energy conditions   [see comment ${\rm i)}$ in Section \ref{Comments}].
Subsequently, we have analysed the properties of perturbations in our model, focusing on the evolution of the effective speed of sound  and that of the Jeans scale. In general, in building a phenomenological model, we have chosen its parameter values  in order to always satisfy the condition  $k \ll k_{\rm J}$ for all $k$ of cosmological interests to which linear theory applies. In this way, we have been able to set theoretical constraints on the parameters of the model, predicting sufficient conditions for the model to be viable.
In particular, we have found that for sufficiently small $c_{\rm s}$, {\it i)}  the Jeans wave number becomes larger and {\it ii)} it becomes vanishingly small for extremely short times, so that the effects caused by its  vanishing  are sufficiently negligible,  as we have showed in  Section \ref{sec:grav-pot} when we have analysed the gravitational potential and the UDM density contrast $\delta = \delta\rho/\rho_{\rm A}$,  Eq.\  (\ref{delta}).
 
Studying observational constraints on  our UDM models  from SN $1$A, CMB 
anisotropies and from the formation of the large-scale structure in the Universe will be the subject of a future 
analysis. In particular, it will be interesting to see if the feature $w_{\rm DE} <-1$ our models show after the transition will be compatible with observations. 
On the theoretical side it will be important to check  if, in a Universe filled with the scalar field component defined by the Lagrangian (\ref{LphiX}), the dynamics will display  the desired behaviour,  with the scalar field mimicking both the DM and DE components.  
Another issue concerning the dynamics of our UDM  is  the possible development of caustics in the non-linear regime.
For instance,  it is well known that inhomogeneous tachyon matter fluctuations could develop caustics, e.g. see \cite{Felder:2002sv, Goswami:2010rs}. However, comparing with these models  our Lagrangian  has an additive potential $V(\phi)$.  These open theoretical issues will be analysed  in  a forthcoming work.

\acknowledgments{DB would like to acknowledge the ICG (Portsmouth) for the hospitality during the development of this project and ``Fondazione Ing. Aldo Gini" for support. DB research has been  partly supported by ASI  contract I/016/07/0 ``COFIS". MB is supported by STFC grant ST/H002774/1. OFP research has been supported by the CNPq contract 150143/2010-9.  Part of the research of DP was carried out at the Jet Propulsion Laboratory, California Institute of Technology, under a contract with the National Aeronautics and Space Administration. The authors also thank  N.\ Bartolo, B. R.\ Crittenden, R.\ Maartens, S.\ Matarrese for discussions and suggestions.}
\appendix

\section{Explicit reconstruction of Lagrangian $\mathcal{L}(\phi,Y)$ }\label{A}

Following the prescriptions in Section \ref{prescription}, the analytic expression of the scalar field Lagrangian for our  UDM model turns out to be
\begin{equation}\label{LphiX}
\mathcal{L}(\phi,Y)=-f(\phi)\sqrt{1-2\frac{h(\phi)}{\left(1+\frac{3}{4}\frac{\phi^2}{1-c_\infty^2}\right)}Y} -V(\phi)\;.
\end{equation}
Now, using the parametric equation of state $p(N)$-$\rho(N)$ defined by  Eqs.\  (\ref{p2})-(\ref{rho2}) and the speed of sound defined in Eq.\ (\ref{cs}), we can infer a relation among the parameters $\mu$, $\nu$, $c_\infty$ and $\beta$, i.e.
\begin{equation}
\mu=\beta\nu(1-c_\infty^2)\;.
\end{equation}
Thanks to this relation we can eliminate $\mu$, finally  obtaining the following potentials
\begin{eqnarray}\label{fphi}
f(\phi) &=&\frac{c_\infty}{\left[1+\left(\frac{4}{3}\frac{1}{c_\infty^2\phi^2}\right)^n\right]^{1/2}}\left\{1-\frac{c_\infty^2}{\left[1+\left(\frac{4}{3}\frac{1}{c_\infty^2\phi^2}\right)^n\right]}\right\}^{-1}\nonumber\\
& &  \left\{ -\frac{\rho_\Lambda}{2}\tanh\left[\frac{\beta \nu}{4}\left(\phi^2-\phi_t^2\right)\right]+2\frac{\rho_\Lambda}{\beta\nu}\frac{1}{\phi^2}\ln\cosh\left[\frac{\beta \nu}{4}\left(\phi^2-\phi_t^2\right)\right]+\frac{4}{3}\frac{\rho_{\rm DE}}{\phi^2}\right\}  \;, \label{fphi} \nonumber \\ \\
 V(\phi) &=& \left\{1-\frac{c_\infty^2}{\left[1+\left(\frac{4}{3}\frac{1}{c_\infty^2\phi^2}\right)^n\right]}\right\}^{-1}\Bigg\{\frac{\rho_\Lambda}{2}\left\{1+\tanh\left[\frac{\beta \nu}{4}\left(\phi^2-\phi_t^2\right)\right]\right\}\nonumber\\
& & \left.-\frac{c_\infty^2}{\left[1+\left(\frac{4}{3}\frac{1}{c_\infty^2\phi^2}\right)^n\right]}\left[\frac{\rho_\Lambda}{2}+2\frac{\rho_\Lambda}{\beta\nu}\frac{1}{\phi^2}\ln\cosh\left[\frac{\beta \nu}{4}\left(\phi^2-\phi_t^2\right)\right]+\frac{4}{3}\frac{\rho_{\rm DE}}{\phi^2}\right]\right\} \;,\label{Vphi}\\
 h(\phi) &=& \frac{\left[1+\frac{1}{1-c_\infty^2}\left(\frac{4}{3}\frac{1}{c_\infty^2\phi^2}\right)^n\right]\left(1+\frac{4}{3}\frac{1-c_\infty^2}{\phi^2}\right)}{\left\{\frac{\rho_\Lambda}{2}+2\frac{\rho_\Lambda}{\beta\nu}\frac{1}{\phi^2}\ln\cosh\left[\frac{\beta \nu}{4}\left(\phi^2-\phi_t^2\right)\right]+\frac{4}{3}\frac{\rho_{\rm DE}}{\phi^2}\right\}\left[1+\left(\frac{4}{3}\frac{1}{c_\infty^2\phi^2}\right)^n\right]} \label{hphi} \;,
\end{eqnarray}
where $\phi_{\rm t}=\left[2/\left(3\nu a_t^{-3}\right)\right]^{1/2}$.

\bibliographystyle{JHEP}
\bibliography{BFastUDM-scalar_field-2}

\providecommand{\href}[2]{#2}\begingroup\raggedright\begin{thebibliography}{10}

\bibitem{Peebles:1984ge}
P.~J.~E. Peebles, {\it {Tests of Cosmological Models Constrained by
  Inflation}},  {\em Astrophys. J.} {\bf 284} (1984) 439--444.

\bibitem{Efstathiou:1990xe}
G.~Efstathiou, W.~J. Sutherland, and S.~J. Maddox, {\it {The cosmological
  constant and cold dark matter}},  {\em Nature} {\bf 348} (1990) 705--707.

\bibitem{Spergel:2003cb}
{\bf WMAP} Collaboration, D.~N. Spergel {\em et~al.}, {\it {First Year
  Wilkinson Microwave Anisotropy Probe (WMAP) Observations: Determination of
  Cosmological Parameters}},  {\em Astrophys. J. Suppl.} {\bf 148} (2003)
  175--194, [\href{http://arxiv.org/abs/astro-ph/0302209}{{\tt
  astro-ph/0302209}}].

\bibitem{Tegmark:2003ud}
{\bf SDSS} Collaboration, M.~Tegmark {\em et~al.}, {\it {Cosmological
  parameters from SDSS and WMAP}},  {\em Phys. Rev.} {\bf D69} (2004) 103501,
  [\href{http://arxiv.org/abs/astro-ph/0310723}{{\tt astro-ph/0310723}}].

\bibitem{Perlmutter:1998np}
{\bf Supernova Cosmology Project} Collaboration, S.~Perlmutter {\em et~al.},
  {\it {Measurements of Omega and Lambda from 42 High-Redshift Supernovae}},
  {\em Astrophys. J.} {\bf 517} (1999) 565--586,
  [\href{http://arxiv.org/abs/astro-ph/9812133}{{\tt astro-ph/9812133}}].

\bibitem{Riess:1998cb}
{\bf Supernova Search Team} Collaboration, A.~G. Riess {\em et~al.}, {\it
  {Observational Evidence from Supernovae for an Accelerating Universe and a
  Cosmological Constant}},  {\em Astron. J.} {\bf 116} (1998) 1009--1038,
  [\href{http://arxiv.org/abs/astro-ph/9805201}{{\tt astro-ph/9805201}}].

\bibitem{Riess:1998dv}
A.~G. Riess {\em et~al.}, {\it {BVRI Light Curves for 22 Type Ia Supernovae}},
  {\em Astron. J.} {\bf 117} (1999) 707--724,
  [\href{http://arxiv.org/abs/astro-ph/9810291}{{\tt astro-ph/9810291}}].

\bibitem{Percival:2007yw}
W.~J. Percival {\em et~al.}, {\it {Measuring the Baryon Acoustic Oscillation
  scale using the SDSS and 2dFGRS}},  {\em Mon. Not. Roy. Astron. Soc.} {\bf
  381} (2007) 1053--1066, [\href{http://arxiv.org/abs/0705.3323}{{\tt
  arXiv:0705.3323}}].

\bibitem{Percival:2009xn}
W.~J. Percival {\em et~al.}, {\it {Baryon Acoustic Oscillations in the Sloan
  Digital Sky Survey Data Release 7 Galaxy Sample}},  {\em Mon. Not. Roy.
  Astron. Soc.} {\bf 401} (2010) 2148--2168,
  [\href{http://arxiv.org/abs/0907.1660}{{\tt arXiv:0907.1660}}].

\bibitem{Amanullah:2010vv}
R.~Amanullah {\em et~al.}, {\it {Spectra and Light Curves of Six Type Ia
  Supernovae at 0.511 < z < 1.12 and the Union2 Compilation}},  {\em Astrophys.
  J.} {\bf 716} (2010) 712--738, [\href{http://arxiv.org/abs/1004.1711}{{\tt
  arXiv:1004.1711}}].

\bibitem{Weinberg:1989}
S.~{Weinberg}, {\it {The cosmological constant problem}},  {\em Reviews of
  Modern Physics} {\bf 61} (Jan., 1989) 1--23.

\bibitem{Zlatev:1998tr}
I.~Zlatev, L.-M. Wang, and P.~J. Steinhardt, {\it {Quintessence, Cosmic
  Coincidence, and the Cosmological Constant}},  {\em Phys. Rev. Lett.} {\bf
  82} (1999) 896--899, [\href{http://arxiv.org/abs/astro-ph/9807002}{{\tt
  astro-ph/9807002}}].

\bibitem{Sahni:1999gb}
V.~Sahni and A.~A. Starobinsky, {\it {The Case for a Positive Cosmological
  Lambda-term}},  {\em Int. J. Mod. Phys.} {\bf D9} (2000) 373--444,
  [\href{http://arxiv.org/abs/astro-ph/9904398}{{\tt astro-ph/9904398}}].

\bibitem{Peebles:2002gy}
P.~J.~E. Peebles and B.~Ratra, {\it {The cosmological constant and dark
  energy}},  {\em Rev. Mod. Phys.} {\bf 75} (2003) 559--606,
  [\href{http://arxiv.org/abs/astro-ph/0207347}{{\tt astro-ph/0207347}}].

\bibitem{Padmanabhan:2002ji}
T.~Padmanabhan, {\it {Cosmological constant: The weight of the vacuum}},  {\em
  Phys. Rept.} {\bf 380} (2003) 235--320,
  [\href{http://arxiv.org/abs/hep-th/0212290}{{\tt hep-th/0212290}}].

\bibitem{Copeland:2006wr}
E.~J. Copeland, M.~Sami, and S.~Tsujikawa, {\it {Dynamics of dark energy}},
  {\em Int. J. Mod. Phys.} {\bf D15} (2006) 1753--1936,
  [\href{http://arxiv.org/abs/hep-th/0603057}{{\tt hep-th/0603057}}].

\bibitem{Tsujikawa:2010sc}
S.~Tsujikawa, {\it {Dark energy: investigation and modeling}},
  \href{http://arxiv.org/abs/1004.1493}{{\tt arXiv:1004.1493}}.

\bibitem{Amendola:1272934}
L.~Amendola and S.~Tsujikawa, {\em Dark energy: Theory and observations}.
\newblock Cambridge Univ. Press, Cambridge, 2010.

\bibitem{Peacock:2006kj}
J.~A. Peacock {\em et~al.}, {\it {Report by the ESA-ESO Working Group on
  Fundamental Cosmology}},  \href{http://arxiv.org/abs/astro-ph/0610906}{{\tt
  astro-ph/0610906}}.

\bibitem{Allen:2004cd}
S.~W. Allen, R.~W. Schmidt, H.~Ebeling, A.~C. Fabian, and L.~van Speybroeck,
  {\it {Constraints on dark energy from Chandra observations of the largest
  relaxed galaxy clusters}},  {\em Mon. Not. Roy. Astron. Soc.} {\bf 353}
  (2004) 457, [\href{http://arxiv.org/abs/astro-ph/0405340}{{\tt
  astro-ph/0405340}}].

\bibitem{Tegmark:2006az}
{\bf SDSS} Collaboration, M.~Tegmark {\em et~al.}, {\it {Cosmological
  Constraints from the SDSS Luminous Red Galaxies}},  {\em Phys. Rev.} {\bf
  D74} (2006) 123507, [\href{http://arxiv.org/abs/astro-ph/0608632}{{\tt
  astro-ph/0608632}}].

\bibitem{Percival:2006kh}
W.~J. Percival, {\it {Cosmological constraints from galaxy clustering}},  {\em
  Lect. Notes Phys.} {\bf 720} (2007) 157--186,
  [\href{http://arxiv.org/abs/astro-ph/0601538}{{\tt astro-ph/0601538}}].

\bibitem{Reid:2009xm}
B.~A. Reid {\em et~al.}, {\it {Cosmological Constraints from the Clustering of
  the Sloan Digital Sky Survey DR7 Luminous Red Galaxies}},  {\em Mon. Not.
  Roy. Astron. Soc.} {\bf 404} (2010) 60--85,
  [\href{http://arxiv.org/abs/0907.1659}{{\tt arXiv:0907.1659}}].

\bibitem{Larson:2010gs}
D.~Larson {\em et~al.}, {\it {Seven-Year Wilkinson Microwave Anisotropy Probe
  (WMAP) Observations: Power Spectra and WMAP-Derived Parameters}},
  \href{http://arxiv.org/abs/1001.4635}{{\tt arXiv:1001.4635}}.

\bibitem{Komatsu:2010fb}
E.~Komatsu {\em et~al.}, {\it {Seven-Year Wilkinson Microwave Anisotropy Probe
  (WMAP) Observations: Cosmological Interpretation}},
  \href{http://arxiv.org/abs/1001.4538}{{\tt arXiv:1001.4538}}.

\bibitem{Blanchard:2010gv}
A.~Blanchard, {\it {Evidence for the Fifth Element Astrophysical status of Dark
  Energy}},  {\em Astron. Astrophys. Rev.} {\bf 18} (2010) 595--645,
  [\href{http://arxiv.org/abs/1005.3765}{{\tt arXiv:1005.3765}}].

\bibitem{Kamenshchik:2001cp}
A.~Y. Kamenshchik, U.~Moschella, and V.~Pasquier, {\it {An alternative to
  quintessence}},  {\em Phys. Lett.} {\bf B511} (2001) 265--268,
  [\href{http://arxiv.org/abs/gr-qc/0103004}{{\tt gr-qc/0103004}}].

\bibitem{Bilic:2001cg}
N.~Bilic, G.~B. Tupper, and R.~D. Viollier, {\it {Unification of dark matter
  and dark energy: The inhomogeneous Chaplygin gas}},  {\em Phys. Lett.} {\bf
  B535} (2002) 17--21, [\href{http://arxiv.org/abs/astro-ph/0111325}{{\tt
  astro-ph/0111325}}].

\bibitem{Bento:2002ps}
M.~C. Bento, O.~Bertolami, and A.~A. Sen, {\it {Generalized Chaplygin gas,
  accelerated expansion and dark energy-matter unification}},  {\em Phys. Rev.}
  {\bf D66} (2002) 043507, [\href{http://arxiv.org/abs/gr-qc/0202064}{{\tt
  gr-qc/0202064}}].

\bibitem{Carturan:2002si}
D.~Carturan and F.~Finelli, {\it {Cosmological Effects of a Class of Fluid Dark
  Energy Models}},  {\em Phys. Rev.} {\bf D68} (2003) 103501,
  [\href{http://arxiv.org/abs/astro-ph/0211626}{{\tt astro-ph/0211626}}].

\bibitem{Sandvik:2002jz}
H.~Sandvik, M.~Tegmark, M.~Zaldarriaga, and I.~Waga, {\it {The end of unified
  dark matter?}},  {\em Phys. Rev.} {\bf D69} (2004) 123524,
  [\href{http://arxiv.org/abs/astro-ph/0212114}{{\tt astro-ph/0212114}}].

\bibitem{Scherrer:2004au}
R.~J. Scherrer, {\it {Purely kinetic k-essence as unified dark matter}},  {\em
  Phys. Rev. Lett.} {\bf 93} (2004) 011301,
  [\href{http://arxiv.org/abs/astro-ph/0402316}{{\tt astro-ph/0402316}}].

\bibitem{Giannakis:2005kr}
D.~Giannakis and W.~Hu, {\it {Kinetic unified dark matter}},  {\em Phys. Rev.}
  {\bf D72} (2005) 063502, [\href{http://arxiv.org/abs/astro-ph/0501423}{{\tt
  astro-ph/0501423}}].

\bibitem{Bertacca:2007ux}
D.~Bertacca, S.~Matarrese, and M.~Pietroni, {\it {Unified dark matter in scalar
  field cosmologies}},  {\em Mod. Phys. Lett.} {\bf A22} (2007) 2893--2907,
  [\href{http://arxiv.org/abs/astro-ph/0703259}{{\tt astro-ph/0703259}}].

\bibitem{Bertacca:2007cv}
D.~Bertacca and N.~Bartolo, {\it {ISW effect in Unified Dark Matter Scalar
  Field Cosmologies: an analytical approach}},  {\em JCAP} {\bf 0711} (2007)
  026, [\href{http://arxiv.org/abs/0707.4247}{{\tt arXiv:0707.4247}}].

\bibitem{Bertacca:2007fc}
D.~Bertacca, N.~Bartolo, and S.~Matarrese, {\it {Halos of Unified Dark Matter
  Scalar Field}},  {\em JCAP} {\bf 0805} (2008) 005,
  [\href{http://arxiv.org/abs/0712.0486}{{\tt arXiv:0712.0486}}].

\bibitem{Balbi:2007mz}
A.~Balbi, M.~Bruni, and C.~Quercellini, {\it {Lambda-alpha DM: Observational
  constraints on unified dark matter with constant speed of sound}},  {\em
  Phys. Rev.} {\bf D76} (2007) 103519,
  [\href{http://arxiv.org/abs/astro-ph/0702423}{{\tt astro-ph/0702423}}].

\bibitem{Quercellini:2007ht}
C.~Quercellini, M.~Bruni, and A.~Balbi, {\it {Affine equation of state from
  quintessence and k-essence fields}},  {\em Class. Quant. Grav.} {\bf 24}
  (2007) 5413--5426, [\href{http://arxiv.org/abs/0706.3667}{{\tt
  arXiv:0706.3667}}].

\bibitem{Pietrobon:2008js}
D.~Pietrobon, A.~Balbi, M.~Bruni, and C.~Quercellini, {\it {Affine
  parameterization of the dark sector: constraints from WMAP5 and SDSS}},  {\em
  Phys. Rev.} {\bf D78} (2008) 083510,
  [\href{http://arxiv.org/abs/0807.5077}{{\tt arXiv:0807.5077}}].

\bibitem{Bertacca:2008uf}
D.~Bertacca, N.~Bartolo, A.~Diaferio, and S.~Matarrese, {\it {How the Scalar
  Field of Unified Dark Matter Models Can Cluster}},  {\em JCAP} {\bf 0810}
  (2008) 023, [\href{http://arxiv.org/abs/0807.1020}{{\tt arXiv:0807.1020}}].

\bibitem{Bilic:2008yr}
N.~Bilic, G.~B. Tupper, and R.~D. Viollier, {\it {Cosmological tachyon
  condensation}},  {\em Phys. Rev.} {\bf D80} (2009) 023515,
  [\href{http://arxiv.org/abs/0809.0375}{{\tt arXiv:0809.0375}}].

\bibitem{Camera:2009uz}
S.~Camera, D.~Bertacca, A.~Diaferio, N.~Bartolo, and S.~Matarrese, {\it {Weak
  lensing signal in Unified Dark Matter models}},  {\em
  Mon.Not.Roy.Astron.Soc.} {\bf 399} (2009) 1995--2003,
  [\href{http://arxiv.org/abs/arXiv:0902.4204}{{\tt arXiv:0902.4204}}]. * Brief
  entry *.

\bibitem{Li:2009mf}
B.~Li and J.~D. Barrow, {\it {Does Bulk Viscosity Create a Viable Unified Dark
  Matter Model?}},  {\em Phys. Rev.} {\bf D79} (2009) 103521,
  [\href{http://arxiv.org/abs/0902.3163}{{\tt arXiv:0902.3163}}].

\bibitem{Piattella:2009kt}
O.~F. Piattella, D.~Bertacca, M.~Bruni, and D.~Pietrobon, {\it {Unified Dark
  Matter models with fast transition}},  {\em JCAP} {\bf 1001} (2010) 014,
  [\href{http://arxiv.org/abs/arXiv:0911.2664}{{\tt arXiv:0911.2664}}].

\bibitem{Gao:2009me}
C.~Gao, M.~Kunz, A.~R. Liddle, and D.~Parkinson, {\it {Unified dark energy and
  dark matter from a scalar field different from quintessence}},  {\em Phys.
  Rev.} {\bf D81} (2010) 043520, [\href{http://arxiv.org/abs/0912.0949}{{\tt
  arXiv:0912.0949}}].

\bibitem{Camera:2010wm}
S.~Camera, T.~D. Kitching, A.~F. Heavens, D.~Bertacca, and A.~Diaferio, {\it
  {Measuring Unified Dark Matter with 3D cosmic shear}},
  \href{http://arxiv.org/abs/1002.4740}{{\tt arXiv:1002.4740}}.

\bibitem{Lim:2010yk}
E.~A. Lim, I.~Sawicki, and A.~Vikman, {\it {Dust of Dark Energy}},  {\em JCAP}
  {\bf 1005} (2010) 012, [\href{http://arxiv.org/abs/1003.5751}{{\tt
  arXiv:1003.5751}}].

\bibitem{Bertacca:2010ct}
D.~Bertacca, N.~Bartolo, and S.~Matarrese, {\it {Unified Dark Matter Scalar
  Field Models}},  \href{http://arxiv.org/abs/1008.0614}{{\tt
  arXiv:1008.0614}}.

\bibitem{DiezTejedor:2006qh}
A.~Diez-Tejedor and A.~Feinstein, {\it {The homogeneous scalar field and the
  wet dark sides of the universe}},  {\em Phys. Rev.} {\bf D74} (2006) 023530,
  [\href{http://arxiv.org/abs/gr-qc/0604031}{{\tt gr-qc/0604031}}].

\bibitem{Brown:1992kc}
J.~D. Brown, {\it {Action functionals for relativistic perfect fluids}},  {\em
  Class. Quant. Grav.} {\bf 10} (1993) 1579--1606,
  [\href{http://arxiv.org/abs/gr-qc/9304026}{{\tt gr-qc/9304026}}].

\bibitem{DiezTejedor:2005fz}
A.~Diez-Tejedor and A.~Feinstein, {\it {Relativistic hydrodynamics with sources
  for cosmological K-fluids}},  {\em Int. J. Mod. Phys.} {\bf D14} (2005)
  1561--1576, [\href{http://arxiv.org/abs/gr-qc/0501101}{{\tt gr-qc/0501101}}].

\bibitem{ArmendarizPicon:1999rj}
C.~Armendariz-Picon, T.~Damour, and V.~F. Mukhanov, {\it {k-Inflation}},  {\em
  Phys. Lett.} {\bf B458} (1999) 209--218,
  [\href{http://arxiv.org/abs/hep-th/9904075}{{\tt hep-th/9904075}}].

\bibitem{Garriga:1999vw}
J.~Garriga and V.~F. Mukhanov, {\it {Perturbations in k-inflation}},  {\em
  Phys. Lett.} {\bf B458} (1999) 219--225,
  [\href{http://arxiv.org/abs/hep-th/9904176}{{\tt hep-th/9904176}}].

\bibitem{Chiba:1999ka}
T.~Chiba, T.~Okabe, and M.~Yamaguchi, {\it {Kinetically driven quintessence}},
  {\em Phys. Rev.} {\bf D62} (2000) 023511,
  [\href{http://arxiv.org/abs/astro-ph/9912463}{{\tt astro-ph/9912463}}].

\bibitem{dePutter:2007ny}
R.~de~Putter and E.~V. Linder, {\it {Kinetic k-essence and Quintessence}},
  {\em Astropart. Phys.} {\bf 28} (2007) 263--272,
  [\href{http://arxiv.org/abs/0705.0400}{{\tt arXiv:0705.0400}}].

\bibitem{Linder:2008ya}
E.~V. Linder and R.~J. Scherrer, {\it {Aetherizing Lambda: Barotropic Fluids as
  Dark Energy}},  {\em Phys. Rev.} {\bf D80} (2009) 023008,
  [\href{http://arxiv.org/abs/0811.2797}{{\tt arXiv:0811.2797}}].

\bibitem{ArmendarizPicon:2000dh}
C.~Armendariz-Picon, V.~F. Mukhanov, and P.~J. Steinhardt, {\it {A dynamical
  solution to the problem of a small cosmological constant and late-time cosmic
  acceleration}},  {\em Phys. Rev. Lett.} {\bf 85} (2000) 4438--4441,
  [\href{http://arxiv.org/abs/astro-ph/0004134}{{\tt astro-ph/0004134}}].

\bibitem{ArmendarizPicon:2000ah}
C.~Armendariz-Picon, V.~F. Mukhanov, and P.~J. Steinhardt, {\it {Essentials of
  k-essence}},  {\em Phys. Rev.} {\bf D63} (2001) 103510,
  [\href{http://arxiv.org/abs/astro-ph/0006373}{{\tt astro-ph/0006373}}].

\bibitem{Vikman:2004dc}
A.~Vikman, {\it {Can dark energy evolve to the phantom?}},  {\em Phys. Rev.}
  {\bf D71} (2005) 023515, [\href{http://arxiv.org/abs/astro-ph/0407107}{{\tt
  astro-ph/0407107}}].

\bibitem{Rendall:2005fv}
A.~D. Rendall, {\it {Dynamics of k-essence}},  {\em Class. Quant. Grav.} {\bf
  23} (2006) 1557--1570, [\href{http://arxiv.org/abs/gr-qc/0511158}{{\tt
  gr-qc/0511158}}].

\bibitem{Babichev:2007dw}
E.~Babichev, V.~Mukhanov, and A.~Vikman, {\it {k-Essence, superluminal
  propagation, causality and emergent geometry}},  {\em JHEP} {\bf 02} (2008)
  101, [\href{http://arxiv.org/abs/0708.0561}{{\tt arXiv:0708.0561}}].

\bibitem{Arroja:2010wy}
F.~Arroja and M.~Sasaki, {\it {A note on the equivalence of a barotropic
  perfect fluid with a K-essence scalar field}},  {\em Phys. Rev.} {\bf D81}
  (2010) 107301, [\href{http://arxiv.org/abs/1002.1376}{{\tt
  arXiv:1002.1376}}].

\bibitem{Unnikrishnan:2010ag}
S.~Unnikrishnan and L.~Sriramkumar, {\it {A note on perfect scalar fields}},
  {\em Phys. Rev.} {\bf D81} (2010) 103511,
  [\href{http://arxiv.org/abs/1002.0820}{{\tt arXiv:1002.0820}}].

\bibitem{Bardeen:1983qw}
J.~M. Bardeen, P.~J. Steinhardt, and M.~S. Turner, {\it {Spontaneous Creation
  of Almost Scale - Free Density Perturbations in an Inflationary Universe}},
  {\em Phys.Rev.} {\bf D28} (1983) 679.

\bibitem{Bruni:1991kb}
M.~Bruni, G.~F.~R. Ellis, and P.~K.~S. Dunsby, {\it {Gauge invariant
  perturbations in a scalar field dominated universe}},  {\em Class. Quant.
  Grav.} {\bf 9} (1992) 921--946.

\bibitem{Mukhanov:2005sc}
V.~Mukhanov, {\it {Physical foundations of cosmology}}, . Cambridge, UK: Univ.
  Pr. (2005) 421 p.

\bibitem{Hu:1998kj}
W.~Hu, {\it {Structure Formation with Generalized Dark Matter}},  {\em
  Astrophys. J.} {\bf 506} (1998) 485--494,
  [\href{http://arxiv.org/abs/astro-ph/9801234}{{\tt astro-ph/9801234}}].

\bibitem{Bardeen:1980kt}
J.~M. Bardeen, {\it {Gauge Invariant Cosmological Perturbations}},  {\em Phys.
  Rev.} {\bf D22} (1980) 1882--1905.

\bibitem{Kodama:1985bj}
H.~Kodama and M.~Sasaki, {\it {Cosmological Perturbation Theory}},  {\em Prog.
  Theor. Phys. Suppl.} {\bf 78} (1984) 1--166.

\bibitem{Piattella:2009da}
O.~F. Piattella, {\it {The extreme limit of the generalized Chaplygin gas}},
  {\em JCAP} {\bf 1003} (2010) 012, [\href{http://arxiv.org/abs/0906.4430}{{\tt
  arXiv:0906.4430}}].

\bibitem{Bruni:1992dg}
M.~Bruni, P.~K.~S. Dunsby, and G.~F.~R. Ellis, {\it {Cosmological perturbations
  and the physical meaning of gauge invariant variables}},  {\em Astrophys. J.}
  {\bf 395} (1992) 34--53.

\bibitem{Mukhanov:1990me}
V.~F. Mukhanov, H.~A. Feldman, and R.~H. Brandenberger, {\it {Theory of
  cosmological perturbations. Part 1. Classical perturbations. Part 2. Quantum
  theory of perturbations. Part 3. Extensions}},  {\em Phys. Rept.} {\bf 215}
  (1992) 203--333.

\bibitem{Visser:1997qk}
M.~Visser, {\it {Energy conditions in the epoch of galaxy formation}},  {\em
  Science} {\bf 276} (1997) 88--90.

\bibitem{Ananda:2005xp}
K.~N. Ananda and M.~Bruni, {\it {Cosmo-dynamics and dark energy with non-linear
  equation of state: A quadratic model}},  {\em Phys. Rev.} {\bf D74} (2006)
  023523, [\href{http://arxiv.org/abs/astro-ph/0512224}{{\tt
  astro-ph/0512224}}].

\bibitem{Ananda:2006gf}
K.~N. Ananda and M.~Bruni, {\it {Cosmo-dynamics and dark energy with a
  quadratic EoS: Anisotropic models, large-scale perturbations and cosmological
  singularities}},  {\em Phys. Rev.} {\bf D74} (2006) 023524,
  [\href{http://arxiv.org/abs/gr-qc/0603131}{{\tt gr-qc/0603131}}].

\bibitem{Bruni-Lazcoz}
{Marco Bruni, Ruth Lazcoz}, {\it {In prep.}}, .

\bibitem{Dodelson:2003ft}
S.~Dodelson, {\it {Modern cosmology}}, . Amsterdam, Netherlands: Academic Pr.
  (2003) 440 p.

\bibitem{Felder:2002sv}
G.~N. Felder, L.~Kofman, and A.~Starobinsky, {\it {Caustics in tachyon matter
  and other Born-Infeld scalars}},  {\em JHEP} {\bf 09} (2002) 026,
  [\href{http://arxiv.org/abs/hep-th/0208019}{{\tt hep-th/0208019}}].

\bibitem{Goswami:2010rs}
U.~D. Goswami, H.~Nandan, and M.~Sami, {\it {Formation of caustics in
  Dirac-Born-Infeld type scalar field systems}},  {\em Phys. Rev.} {\bf D82}
  (2010) 103530, [\href{http://arxiv.org/abs/1006.3659}{{\tt
  arXiv:1006.3659}}].

\end{thebibliography}\endgroup
\end{document}